\documentclass[11pt]{article}%
\usepackage{amsfonts}
\usepackage{float}
\usepackage{amssymb}
\usepackage{graphicx}
\usepackage{setspace}
\usepackage[round]{natbib}
\usepackage{amsmath}%
\usepackage{amsthm}%
\usepackage{palatino}
\usepackage[top=8pc,bottom=8pc,left=8pc,right=8pc]{geometry}

\title{Sequential Bayesian Analysis of Multivariate Count Data}

\author{Tevfik Aktekin\\
\textit{Decision Sciences}\\
\textit{University of New Hampshire}\footnote{Aktekin is an associate professor of decision sciences at the Peter T. Paul College of Business and Economics in the University of New Hampshire. email: tevfik.aktekin@unh.edu. Polson is a Professor of Econometrics and Statistics
at the Chicago Booth School of Business. email: ngp@chicagobooth.edu. Soyer is a professor of decision sciences at the George Washington University School of Business. email: soyer@gwu.edu. }\\
\\
Nicholas G. Polson\\
\textit{Booth School of Business}\\
\textit{University of Chicago}\\
\\
Refik Soyer\\
\textit{Decision Sciences}\\
\textit{George Washington University}\\
\\
}

\date{}

\begin{document}

\maketitle
\begin{abstract}
\noindent We develop a new class of dynamic multivariate Poisson count models that allow for fast online updating and we refer to these models as multivariate Poisson-scaled beta (MPSB). The MPSB model allows for serial dependence in the counts as well as dependence across multiple series with a random common environment. Other notable features include analytic forms for state propagation and predictive likelihood densities. Sequential updating occurs through the updating of the sufficient statistics for static model parameters, leading to a fully adapted particle learning algorithm and a new class of predictive likelihoods and marginal distributions which we refer to as the (dynamic) multivariate confluent hyper-geometric negative binomial distribution (MCHG-NB) and the the dynamic multivariate negative binomial (DMNB) distribution. To illustrate our methodology, we use various simulation studies and count data on weekly non-durable goods consumer demand.
\vspace{0.5pc}

\end{abstract}
\newpage

\section{Introduction}
Data on discrete valued counts pose a number of statistical modeling challenges despite their widespread applications in web analytics, epidemiology, economics, finance, operations, and other fields. For instance, Amazon, Facebook and Google often are interested in modeling and predicting the number of (virtual) customer arrivals during a specific time period or policy makers require predicting the number of individuals who possess a common trait for resource deployment and allocation purposes. In online settings, the challenge then is fast and efficient prediction of web trafficking counts from multiple websites and pages over time. The total number of clicks over time may be positively dependent with the counts the main site receives and there is a need for dynamic multivariate count models. Thus, we develop a dynamic (state-space) multivariate Poisson model together with particle filtering and learning methods for sequential online updating \citep{GSS93, CJLP10}. We account for dependence over time and across series, via a scaled beta state evolution and a random common environment. Our model is termed the multivariate Poisson-scaled beta (MPSB). As a by-product, we introduce two new multivariate distributions, the dynamic multivariate negative binomial (DMNB) and the multivariate confluent hyper-geometric negative binomial (MCHG-NB) distributions which correspond to marginal and predictive distributions.    

Recent advances in discrete valued time series can be found in \cite{Nbook15}. However, there is little work on count data models which accounts for serial dependence. Typically, the dependence between time series of counts can be modeled either using traditional stationary time series models \citep{AA87, Z88, FM04} which are known as \textit{observation driven} models \citep{Cox81} or via state space models \citep{HF89, DK00, FW06, AS11,ASX13, SGF13} that are known as \textit{parameter driven} models. In a state space model, the dependence between the counts is captured via latent factors who follow some form of a stochastic process. These type of models generally assume conditional independence of the counts given the latent factors as opposed to stationary models where counts are always unconditionally dependent.    

Analysis of discrete valued multivariate time series has so far been limited due to computational challenges. In particular, little attention has been given to multivariate models and our approach is an attempt to fill this gap. For example, \cite{PK11, PK12} use observation driven, more specifically multivariate INAR(1) models. \cite{Nalini14} uses Bayesian observation driven models and introduces a hierarcical multivariate Poisson time series model. Markov chain Monte Carlo (MCMC) methods are used for computation where the evaluation of the multivariate Poisson likelihood requires a significant computational effort. \cite{Nalini15a} develops zero-inflated Poisson models for multivariate time series of counts and \cite{Nalini15b} study finite mixtures of multivariate Poisson time series. State-space models of multivariate count data was presented in \cite{Ord93} and in \cite{J99} using the EM algorithm. Closely related models of correlated Poisson counts in a temporal setting include research on marked Poisson processes as in \cite{T10, T12, DHDC12}.    

One advantage of parameter driven models is that the previous correlations are captured by time evolution of the state parameter which we refer to as the \textit{random common environment}. The correlations among the multiple series are induced by this random common environment that follows a Markovian evolution, as in \cite{SM86,ASX13,SGF13}, and modulates the behavior of individual series. The idea of the random common environment is widely used in risk analysis \citep{AK51} and reliability \citep{LS86} literatures to model dependence. Our strategy of using the random common environment provides a new class of models for multivariate counts  that can be considered to be dynamic extensions of models considered in \cite{AK51}.    

Sequential Bayesian analysis (\cite{PSM08,CJLP10}) and forecasting requires the use of sequential Monte Carlo techniques. MCMC methods via the forward filtering backward sampling (FFBS) of \cite{CK94} and \cite{FS94} are not computationally efficient since it requires rerunning of chains to obtain filtering distributions with each additional observation. Particle filtering (PF) and particle learning (PL) methods avoid this computational burden to esimate the dynamic state as well as the static parameters in an efficient manner. As pointed out by \cite{CJLP10}, estimating static parameters within the PF framework is notoriously difficult especially in higher dimensions. However, given the specific structure of the proposed state space model (as the conditional filtering densities of all the static parameters can be obtained in closed form with conditional sufficient statistics), it is possible to develop such a filtering scheme that can be used for both on-line updating and forecasting. 

The rest of the paper is organized as follows. Section 2 introduces our multivariate time series model for counts and develops its properties. Section 3 briefly reviews some of the PF and PL methods with a focus on Poisson count data. The proposed model and estimation algorithms are illustrated in Section 4 using calibration studies and an actual data set on weekly time series of consumer demand for non-durable goods. Section 5 provides concluding remarks, discussion of limitations and future work. 

\section{Multivariate Poisson-Scaled Beta (MPSB) Model}
Suppose that we observe $\lbrace (Y_{11}, \ldots, Y_{1T}),\ldots, (Y_{J1}, \ldots, Y_{JT}) \rbrace$, a sequence of evenly spaced counts observed up until time $T$ for $J$ series. We assume that these $J$ series are exposed to the same external environment similar to the common operational conditions for the components of a system as considered by
\cite{LS86} in reliability analysis. The analysis of financial and economic time series also includes several series that are affected by the same economic swings in the market. To account for such dependence, we assume a Bayesian hierarchical model of the form  
\begin{equation}
(Y_{jt}|\lambda_{j}, \theta_{t}) \sim Pois(\lambda_{j}\theta_{t}), \: \text{for} \: j=1,\ldots, J \: \text{and} \: t=1,\ldots, T,  
\label{multimod}
\end{equation}
where $\lambda_j$ is the rate specific to the $j$th series and $\theta_t$ represents the effects of the random common environment modulating $\lambda_j$. 
Following \cite{SM86}, a Markovian evolution is assumed for $\theta_t$ as 
\begin{equation}
\theta_{t}= \frac{\theta_{t-1}}{\gamma} \epsilon_{t},
\label{multimod2}
\end{equation}
where the error terms follow a Beta distribution as,  
\begin{equation*}
(\epsilon_{t}|D^{t-1},\lambda_{1},\ldots, \lambda_{J}) \sim Beta[\gamma \alpha_{t-1}, (1-\gamma) \alpha_{t-1}], 
\end{equation*}
where $\alpha_{t-1} > 0, 0 < \gamma < 1$ and $D^{t-1} = \lbrace D^{t-2}, Y_{1,t-1}, \ldots, Y_{J,t-1} \rbrace$ represents the sequential arrival of data. We refer to this class of models as multivariate Poisson-scaled beta (MPSB) models due to the relationship between the observation and state equations. We also note here that the state equation above (as discussed in \cite{SM86}) is defined conditional on previous counts unlike the state equations in traditional dynamic linear models. 

\subsection{Dynamic Online Bayesian Updating}
The observation model (\ref{multimod}), is a function of both the dynamic environment $\theta_{t}$ and the static parameters, $\lambda_{j}$'s. For example, in the case where $Y_{jt}$ represents the weekly consumer demand for household $j$ at time $t$, $\lambda_j$ accounts for the effects of the household specific rate and $\theta_t$ for the effects of the random common economic environment that both households are exposed to at time $t$. When $\theta_t>1$, the environment is said to be more favorable than usual which leads to a higher overall Poisson rate and vice versa. 
In the evolution equation (\ref{multimod2}), the term $\gamma$ acts like a discount factor common for all $j$ series. For notational convenience, we suppress the dependence of all conditional distributions on $\gamma$ in our discussion below. Having the state evolution as (\ref{multimod2}) also implies the following scaled beta density for $(\theta_{t}|\theta_{t-1})$ 
\begin{equation}
p(\theta_{t}|\theta_{t-1},D^{t-1},\boldsymbol{\lambda}) = \dfrac{\Gamma(\alpha_{t-1})}{\Gamma(\gamma\alpha_{t-1})\Gamma((1-\gamma)\alpha_{t-1})} \Big( \dfrac{\gamma}{\theta_{t-1}}\Big)^{\gamma\alpha_{t-1}} \theta_{t}^{\gamma\alpha_{t-1}-1} \Big( 1- \dfrac{\gamma}{\theta_{t-1}}\theta_{t} \Big)^{(1-\gamma)\alpha_{t-1}}, 
\label{multimod3}
\end{equation}
where $(\theta_{t}|\theta_{t-1},D^{t-1},\boldsymbol{\lambda})$ is defined over $(0;\dfrac{\theta_{t-1}}{\gamma})$ and the vector of static parameters is defined as $\boldsymbol{\lambda}=\lbrace \lambda_{1}, \ldots, \lambda_{J} \rbrace$. 

Here, we assume that for component $j$, given  $\theta_{t}$'s and $\lambda_j$, $Y_{jt}$'s are conditionally independent over time. Furthermore, we assume that at time $t$, given  $\theta_{t}$ and $\lambda_j$'s, $Y_{jt}$'s are conditionally independent of each other.      

Conditional on the static parameters, it is possible to obtain an analytically tractable filtering of the states. At time 0, prior to observing any count data, we assume that $(\theta_{0}|D^{0}) \sim Gamma(\alpha_{0}, \beta_{0})$, then by induction we can show that 
\begin{equation}
(\theta_{t-1}|D^{t-1},\boldsymbol{\lambda}) \sim Gamma(\alpha_{t-1},\beta_{t-1}), 
\label{priort}
\end{equation}
and using (\ref{multimod3}) and (\ref{priort}) show that the prior for $\theta_t$ would be  
\begin{align}
p(\theta_{t}|D^{t-1},\boldsymbol{\lambda}) & = \int p(\theta_{t}|\theta_{t-1},D^{t-1},\boldsymbol{\lambda})p(\theta_{t-1}|D^{t-1},\boldsymbol{\lambda})d\theta_{t-1} \\
& \sim Gamma(\gamma\alpha_{t-1},\gamma\beta_{t-1}).
\label{conprior}
\end{align}
Therefore, the filtering density at time $t$ can be obtained using (\ref{multimod}) and (\ref{conprior}) as  
\begin{align}
p(\theta_{t}|D^{t},\boldsymbol{\lambda}) & \propto p(Y_{1t},\ldots,Y_{jt}|\theta_{t}, \boldsymbol{\lambda}) p(\theta_{t}|D^{t-1},\boldsymbol{\lambda})\\
& \propto \Bigg (\prod_{j} (\theta_{t}\lambda_{j})^{Y_{jt}} e^{-\lambda_{j}\theta_{t}} \Bigg) \Bigg ( \theta_{t}^{\gamma\alpha_{t-1}-1} e^{- \gamma\beta_{t-1}\theta_{t}} \Bigg), 
\label{filterdens}
\end{align}
which is  
\begin{equation}
(\theta_{t}|D^{t},\boldsymbol{\lambda}) \sim Gamma(\alpha_{t}, \beta_{t}), 
\label{filter2}
\end{equation}
where $\alpha_{t}= \gamma\alpha_{t-1}+ (Y_{1t} + \ldots + Y_{Jt})$ and $\beta_{t}= \gamma \beta_{t-1} + (\lambda_{1} + \ldots + \lambda_{J})$. As a consequence, both the effects of all counts as well as the individual effects of each series are used in updating the random common environment. 
\subsection{Dynamic Multivariate Negative Binomial (DMNB) Distribution}
An important feature of the model is the availability of the marginal distribution of $Y_{jt}$ conditional on $\lambda_{j}$'s for $j=1, \ldots, J$. This is given by 
\begin{align}
p(Y_{jt}|\boldsymbol{\lambda}, D^{t-1}) & = \int p(Y_{jt}|\theta_{t}, \lambda_{j}) p(\theta_{t}|D^{t-1},\boldsymbol{\lambda})  d\theta_{t} \\
&= \binom{\gamma\alpha_{t-1}+Y_{jt}-1}{Y_{jt}} \Big (1-\frac{\lambda_{j}}{\gamma \beta_{t-1}+\lambda_{j}} \Big)^{\gamma\alpha_{t-1}} \Big( \frac{\lambda_{j}}{\gamma \beta_{t-1}+\lambda_{j}} \Big)^{Y_{jt}},
\label{marglike}
\end{align}
which is a negative binomial model denoted as $\text{NB}(\gamma \alpha_{t-1}, \frac{\lambda_{j}}{\gamma \beta_{t-1} + \lambda_{j}})$, where $\frac{\lambda_{j}}{\gamma \beta_{t-1} + \lambda_{j}}$ is the probability of success. From the conditional independence assumptions, we can obtain the multivariate distribution of $\bold{Y_{t}} = \lbrace Y_{1t}, \ldots, Y_{Jt} \rbrace $ conditional on $\lambda_{j}$'s as 
\begin{small}
\begin{equation}
p(\bold{Y_{t}}|\boldsymbol{\lambda}, D^{t-1}) = \frac{\Gamma(\gamma\alpha_{t-1}+ \sum_{j} Y_{jt})}{\Gamma(\gamma\alpha_{t-1}) \prod_{j} \Gamma(Y_{jt}+1)}  \prod_{j} \Bigg (\frac{\lambda_{j}}{\gamma \beta_{t-1}+\sum _{j} \lambda_{j}} \Bigg)^{Y_{jt}} \Bigg( \frac{\gamma \beta_{t-1}}{\gamma \beta_{t-1}+ \sum_{j} \lambda_{j}} \Bigg )^{\gamma\alpha_{t-1}}.
\end{equation}
\end{small}
This is a generalization of the traditional negative binomial distribution. We refer to this distribution as the dynamic multivariate negative binomial (DMNB) distribution which will play an important role in learning about the discount parameter, $\gamma$. Therefore, the bivariate distribution,  $p(Y_{it}, Y_{jt}|\boldsymbol{\lambda}, D^{t-1})$, for  series $i$ and $j$, is given by
\begin{small}
\begin{equation}
\frac{\Gamma(\gamma\alpha_{t-1}+Y_{it}+Y_{jt})}{\Gamma(\gamma\alpha_{t-1})%
\Gamma(Y_{it}+1)\,\Gamma(Y_{jt}+1)\,\,}\,\Bigl(\frac{\gamma\beta_{t-1}\,\,}{\,%
\lambda_i+\lambda_j+\gamma\beta_{t-1}}\Bigr)^{\gamma\alpha_{t-1}}\Bigl(\frac{%
\lambda_i\,\,}{\,\lambda_i+\lambda_j+\gamma\beta_{t-1}}\Bigr)^{Y_{it}}\Bigl(%
\frac{\lambda_j\,\,}{\,\lambda_i+\lambda_j+\gamma\beta_{t-1}}\Bigr)^{Y_{jt}}\,
\label{bivariate}
\end{equation}
\end{small}
which is a bivariate negative binomial distribution with integer values of $\gamma\alpha_{t-1}$. We note that (\ref{bivariate}) is the dynamic version of the negative binomial distribution from \cite{AK51} who considered it for modeling the number of industrial accidents in a workplace such as a production facility. Furthermore, the conditional distributions of $Y_{jt}$'s will also be negative binomial type distributions. The conditional mean, or the regression of $Y_{jt}$ given $Y_{it}$ is a linear function $Y_{it}$ given by  
\begin{equation}
E[Y_{jt}|Y_{it},\boldsymbol{\lambda},
D^{t-1}]=\frac{\lambda_j(\gamma\alpha_{t-1}+Y_{it})}{(\lambda_i+\gamma%
\beta_{t-1})}.
\end{equation}
The bivariate counts are positively correlated with correlation given by 
\begin{equation}
Cor(Y_{it},Y_{jt}|\boldsymbol{\lambda},
D^{t-1})=\sqrt{\frac{\lambda_i\lambda_j}{(\lambda_i+\gamma\beta_{t-1})(%
\lambda_j+\gamma\beta_{t-1})}}.
\label{cor}
\end{equation}
Given (\ref{cor}), our proposed model would be suitable for series that are only positively correlated. One of our examples which will be presented in our numerical illustration section will include counts of weekly demand for consumer non-durable goods of several households that are positively correlated with each other. Also, the structure (\ref{cor}) suggests that as $\gamma$ approaches zero (or very small values), for the same values of $\lambda_{j}$'s, the correlation between  two series increases. A similar argument can be made by observing the state equation (\ref{multimod2}) where $\gamma$ was introduced as a common discount parameter. In our simulations and analysis of real count data, we only consider series that are positively correlated and discuss its implications. Even tough this is a limitation of our model, it is possible to find positively correlated time series of counts in many fields when the series are assumed to be exposed to the same environment.  

\subsection{Forward Filtering and Backward Sampling (FFBS)}
In what follows, we introduce and discuss methods for sequentially estimating the dynamic state parameters, $\theta_{t}$'s, the static parameters, $\lambda_{j}$'s and the discount factor $\gamma$. We first assume that $\gamma$ is known.

We assume that apriori $\lambda_{j}$'s are independent of each other as well as $\theta_{0}$ and having gamma priors as
\begin{equation}
\lambda_{j} \sim Gamma(a_{j}, b_{j}),  \: \text{for} \: j=1,\ldots, J.
\label{gamprior}
\end{equation}

The model can be either estimated using MCMC techniques or particle filtering methods. For MCMC, one needs to generate samples from the joint posterior of all parameters as in $p(\boldsymbol{\theta_{t}}, \boldsymbol{\lambda}|D^{t})$ where $\boldsymbol{\theta_{t}}=\lbrace \theta_{1} \ldots, \theta_{t} \rbrace$ using a Gibbs sampling scheme via the following steps 
\begin{enumerate}
\item Generate $\theta_{t}$'s via $p(\theta_{1},\ldots, \theta_{t}|\lambda_{1}, \ldots, \lambda_{j}, D^{t})$
\item Generate $\lambda_{j}'$s via $p(\lambda_{1},\ldots, \lambda_{j}| \theta_{1},\ldots, \theta_{t}, D^{t})$
\end{enumerate}
In step 1, the forward filtering and backward sampling (FFBS) can  be used to estimate the conditional joint distribution of the state parameters where the joint density $p(\theta_{1},\ldots, \theta_{t}|\boldsymbol{\lambda}, D^{t})$ can be factored as
$$
p(\theta_{t}|\boldsymbol{\lambda}, D^{t}) p(\theta_{t-1}|\theta_{t},\boldsymbol{\lambda}, D^{t-1})\cdots p(\theta_{1}|\theta_{2},\boldsymbol{\lambda}, D^{1}).
$$  
The implementation of FFBS would be straightforward in our model as we have the following shifted gamma densities where $\gamma \theta_{t}<\theta_{t-1}$
$$
(\theta_{t-1}|\theta_{t},\boldsymbol{\lambda}, D^{t-1})\sim Gamma[(1-\gamma) \alpha_{t-1}, \beta_{t-1}].
$$
In Step 2, we can use the Poisson-Gamma conjugacy,
\begin{align*}
p(\lambda_{j}|\boldsymbol{\theta}, D^{t}) & \propto p(Y_{j1},\ldots,Y_{jt}|\theta_{t}, \lambda_{j}) p(\lambda_{j})  \\
& \propto \Bigg (\prod_{t} (\theta_{t}\lambda_{j})^{Y_{jt}} e^{-\lambda_{j}\theta_{t}} \Bigg) \Bigg ( \lambda_{j}^{a_{j}-1} e^{- b_{j}\lambda_{j}} \Bigg),
\label{fullcondlamdacalc}
\end{align*}
which is a gamma density as  
\begin{equation}
(\lambda_{j}|\boldsymbol{\theta_{t}}, D^{t}) \sim Gamma(a_{jt}, b_{jt}),  
\label{fullcondlamda}
\end{equation}
where $a_{jt}= a_{j} + (Y_{j1} + \ldots + Y_{jt})$ and $b_{jt} = b_{j} + (\theta_{1} + \ldots + \theta_{t})$. It is important to observe that given the state parameters, $\boldsymbol{\theta_{t}}$ and data,  $\lambda_{j}$'s are conditionally independent. However, unconditionally they will not necessarily be independent whose implications are investigated in our numerical example. The availability of (\ref{fullcondlamda}) and more importantly the sequential updating of its parameters using sufficient statistics is important in developing particle learning methods which we discuss in detail in the sequel. 

As pointed out by \cite{S02} and \cite{CJLP10}, the issue with MCMC methods in state space models is that the chains need to be restarted for every data point observed and the simulation dimension becomes larger as we observe more data over time. Furthermore, MCMC methods require convergence of chains via the calibration of thinning intervals (to reduce autocorrelation of the samples) and the determination of the burn-in period's size, both of which would increase the computational burden. Therefore, using MCMC methods would not be ideal for sequential updating whose implications we investigate in our numerical example section. However, the FFBS algorithm can be used to obtain smoothing estimates in a very straightforward manner since, unlike filtering, smoothing does not require sequentially restarting the chains. In a single block run of the above FFBS algorithm, one can obtain estimates of $(\theta_{1}, \ldots, \theta_{t}|D^{t})$ by collecting the associated samples generated from $
p(\theta_{1}, \ldots, \theta_{t}, \boldsymbol{\lambda}| D^{t})$. When fast sequential estimation is of interest, an alternative approach is the use of particle filtering (PF) techniques that are based on the idea of re-balancing a finite number of particles of the posterior states given the next data point proportional to its likelihood.

\section{Particle Learning of the MPSB Model}
For sequential state filtering and parameter learning, we make use of the particle learning (PL) method of \cite{CJLP10} to update both the dynamic and the static parameters. To summarize, the PL approach starts with the resampling of state particles at time $t$ using weights proportional to the predictive likelihood which ensures that the highly likely particles are moved forward. The resampling step is followed by the propagation of the current state ($t$) to the future state ($t+1$). Note that in both the resampling and  propagation steps, one-step-ahead observations are used. The last step involves updating the static parameters by computing the conditional sufficient statistics. Even tough there has been several applications of the PL methods in the literature, none of them focus on the analysis of Poisson count data. Among many other successful applications, some recent work of the PL algorithm include \cite{CLPT10} for estimating general mixtures, \cite{GP11} for estimating Gaussian process models in sequential design and optimization, and \cite{LP10} for estimating fat-tailed distributions. 

Let us first assume that $\gamma$ is known and define $z_{t}$ as the essential vector of parameters to keep track of at each $t$. The essential vector will consist of the dynamic state parameter ($\theta_{t}$), static parameters ($\boldsymbol{\lambda}$) and conditional sufficient statistics $s_{t}=f(s_{t-1}, \theta_t,\bold{Y_{t}} )$ for updating the static parameters. The fully adapted version of PL can be summarized as follows using the traditional notation of PF methods     

\begin{enumerate}
\item (Resample) $\lbrace z_{t} \rbrace_{i=1}^{N}$ from $z_{t}^{(i)}= \lbrace s_{t}, \boldsymbol{\lambda} \rbrace^{(i)}$  using weights $w_{t}^{(i)} \propto p(\bold{Y_{t+1}} |z_{t}^{(i)})$ 
\item (Propagate) $\lbrace \theta_{t}^{(i)} \rbrace $ to $\lbrace \theta_{t+1}^{(i)} \rbrace$ via $p(\theta_{t+1}|z_{t}^{(i)}, \bold{Y_{t+1}})$
\item (Update) $s_{t+1}^{(i)} = f(s_{t}^{(i)}, \theta_{t+1}^{(i)},\bold{Y_{t+1}})$ 
\item (Sample) $(\boldsymbol{\lambda})^{(i)}$ from $p(\boldsymbol{\lambda}|s_{t+1}^{(i)})$
\end{enumerate} 

In step 1, note that $z_{t}$ will be stored at each point in time and only includes one state parameter ($\theta_{t}$), hence eliminating the need to update all state parameters ($\boldsymbol{\theta_{t}}$) jointly for each time new data is observed. In step 1, $s_{t}$ represents the ensemble of conditional sufficient statistics for updating the static parameters. From (\ref{fullcondlamda}), it is easy to see that $s_{t}$ should only consist of $\bold{Y_{t}}$ and $\theta_{t}$ if we rewrite $a_{jt} = a_{j,t-1} + Y_{jt}$ and $b_{jt} = b_{j,t-1} + \theta_{t}$ for each $j$. In step 3, $f(.)$ represents this deterministic updating of the conditional sufficient statistic based on the $a_{jt}$ and $b_{jt}$ recursions. 

In order for the above PL scheme to work, we need $p(\bold{Y_{t+1}} |z_{t}^{(i)})$, the predictive likelihood, for computing the weights in step 1 and $p(\theta_{t+1}|z_{t}^{(i)}, \bold{Y_{t+1}})$, the propagation density, for step 2. Note that this propagation density is not the same state evolution equation from (\ref{multimod3}) due to the inclusion of $\bold{Y_{t+1}}$ in the conditioning argument, which ensures that the most current data is considered in propagating the states. Next, we present the summary of how these two quantities can be obtained. A detailed version can be found in Appendix A. We also show in detail the conjugate nature of our model in the Appendix B with the detailed steps of how the dynamic multivariate version was obtained starting with the static univariate model. 

\subsection*{Step 1: Obtaining the resampling weights}
The predictive likelihood is denoted by $p(\bold{Y_{t+1}} |z_{t})=p(\bold{Y_{t+1}} |\theta_{t}, \boldsymbol{\lambda}, D^{t})$ and is required to compute the resampling weights in step 1 of the above PL algorithm. Specifically, we need to compute  
\begin{align*} 
w_{t} = p(\bold{Y_{t+1}} |\theta_{t}, \boldsymbol{\lambda}, D^{t}) = \int p(\bold{Y_{t+1}}|\theta_{t+1},\boldsymbol{\lambda})p(\theta_{t+1}|\theta_{t},\boldsymbol{\lambda}, D^{t}) d\theta_{t+1}, 
\end{align*}
where $p(\bold{Y_{t+1}}|\theta_{t+1},\boldsymbol{\lambda})$ is the product of the Poisson likelihoods (\ref{multimod}) and $p(\theta_{t+1}|\theta_{t},\boldsymbol{\lambda}, D^{t})$ is the state equation (\ref{multimod3}). We can show that $w_{t}$ to be equal to  
\begin{equation} 
w_{t} =\Bigg (\prod_{j}  \frac{\lambda_{j}^{Y_{j,t+1}}}{Y_{j,t+1}!} \Bigg) \Bigg( \frac{\theta_{t}}{\gamma} \Bigg)^{\sum_{j} Y_{j,t+1}} \Bigg( \dfrac{\Gamma(\sum_{j}Y_{j,t+1}+\gamma \alpha_{t} )\Gamma(\alpha_{t})}{\Gamma(\sum_{j}Y_{j,t+1} + \alpha_{t})\Gamma(\gamma \alpha_{t})} \Bigg)CHF(a;a+b;-c),  
\label{weight}
\end{equation}
where $a=\sum_{j}Y_{j,t+1}+\gamma \alpha_{t}, a+b=\sum_{j}Y_{j,t+1} + \alpha_{t}, c= (\sum_{j} \lambda_{j}) \frac{\theta_{t}}{\gamma}$. Here, CHF represents the confluent hyper-geometric function of \cite{AS68}. For evaluating the CHF function, fast computation methods exist; see for instance the \textit{gsl} package in R by \cite{gsl}. The resampling weights (\ref{weight}) also represent the predictive likelihood (marginal) for the proposed class of dynamic multivariate Poisson models. To the best of our knowledge, (\ref{weight}) represents the form of a new multivariate distribution which we refer to as (dynamic) multivariate confluent hyper-geometric negative binomial distribution (MCHG-NB); see the Appendix B for the details. 

\subsection*{Step 2: Obtaining the propagation density}
The propagation density in step 2 of the PL algorithm can be shown to be  
\begin{equation*} 
p(\theta_{t+1}|\theta_{t},\boldsymbol{\lambda},\bold{Y_{t+1}},D^t) \propto \theta_{t+1}^{(\sum_{j}Y_{j,t+1}) + \gamma \alpha_{t} -1}  \Big( 1-  \frac{\gamma}{\theta_{t}}\theta_{t+1} \Big)^{(1-\gamma)\alpha_{t}-1} e^{-(\sum_{j} \lambda_{j}) \theta_{t+1}}.
\end{equation*}
The above form is proportional to a scaled hyper-geometric beta density (see \cite{gordy1998comp}) defined over the range $(0;\frac{\theta_{t}}{\gamma})$, as $HGB(a,b,c)$, with parameters 
$$a=(\sum_{j}Y_{j,t+1}) + \gamma \alpha_{t}, b=(1-\gamma)\alpha_{t} \: \text{and} \: c=\sum_{j}  \lambda_{j}.$$ 
To generate samples from the HGB density, it is possible to use a rejection sampling based approach. First, we can numerically evaluate the maximum of the HGB density over (0,1) using a non-linear numerical search technique and use the maximum as an enveloping constant for developing a rejection sampling algorithm. We comment on the performance of the sampling method in our numerical section and also provide an alternative below.   

Now that we have both the predictive likelihood for computing the resampling weights and the propagation density, the PL algorithm can be summarized as  

\begin{enumerate}
\item (Resample) $\lbrace z_{t} \rbrace_{i=1}^{N}$ from $z_{t}^{(i)}= \lbrace s_{t}, \boldsymbol{\lambda} \rbrace^{(i)}$  using weights 
$$w_{t}^{(i)} \propto \Bigg (\prod_{j}  \frac{\lambda_{j}^{Y_{j,t+1}}}{Y_{j,t+1}!} \Bigg) \Bigg( \frac{\theta_{t}}{\gamma} \Bigg)^{\sum_{j} Y_{j,t+1}} \Bigg( \dfrac{\Gamma(\sum_{j}Y_{j,t+1}+\gamma \alpha_{t} )\Gamma(\alpha_{t})}{\Gamma(\sum_{j}Y_{j,t+1} + \alpha_{t})\Gamma(\gamma \alpha_{t})} \Bigg)CHF(a;a+b;-c)$$
\item (Propagate) $\lbrace \theta_{t}^{(i)} \rbrace $ to $\lbrace \theta_{t+1}^{(i)} \rbrace$ via $HGB[(\sum_{j}Y_{j,t+1}) + \gamma \alpha_{t},(1-\gamma)\alpha_{t}, \sum_{j}  \lambda_{j}]$ defined over $(0;\frac{\theta_{t}}{\gamma})$
\item (Update) $s_{t+1}^{(i)} = f(s_{t}^{(i)}, \theta_{t+1}^{(i)},\bold{Y_{t+1}})$ 
\item (Sample) $(\boldsymbol{\lambda})^{(i)}$ from $p(\lambda_{j}|s_{t+1}^{(i)}) \sim Gamma(a_{j,t+1}, b_{j, t+1}) \: \text{for} \: j=1,\ldots, J$.
\end{enumerate} 
The availability of the recursive updating for the sufficient statistics of the static parameters makes our model an ideal candidate for applying the PL method. Note that in step 4, the conditional distributions of the static parameters are coming from (\ref{fullcondlamda}). Alternatively, if generating from the HGB distribution in step 2 is not computationally efficient, then one can use another step in the vein of sequential importance sampling by resampling the $\theta_{t+1}$'s using weights proportional to the likelihood. For instance, we can replace step 2 in the above with 
\begin{itemize}
\item (Propagate) $\lbrace \tilde{\theta}_{t+1}\rbrace^{(i)}$ from $p(\theta_{t+1}|\theta_{t}^{(i)}, \boldsymbol{\lambda}^{(i)}, D^{t})$
\item (Resample) $\lbrace \theta_{t+1}\rbrace^{(i)}$ using weights $w_{t+1} \propto p(\bold{Y_{t}}|\tilde{\theta}_{t+1}^{(i)}, \boldsymbol{\lambda}^{(i)} )$
\end{itemize}
We comment on the performance of the above approach in our numerical example. 

\subsection*{Updating the discount factor $\gamma$}
For the sequential estimation of the $\gamma$ posterior at each point in time, we make use of the availability of the marginal likelihood conditional on the $\lambda_{j}$'s which is a dynamic multivariate negative binomial density. Estimation of a static parameter that does not evolve over time is surprisingly challenging in a PL context. It is not possible to incorporate the estimation of $\gamma$ in step 5 of the above algorithm using an importance sampling step as it will lead to the well known particle degeneracy issue. Unlike the $\lambda_{j}$'s, the conditional  posterior distribution of $\gamma$ is not a known density with deterministic conditional recursive updating. Therefore, for models where $\gamma$ is treated as an unknown quantity, we suggest the use of the marginal likelihood conditional on the $\lambda_{j}$'s from (\ref{marglike}). Therefore, we can write the conditional posterior of $\gamma$ as 
\begin{equation}
p(\gamma=k|\boldsymbol{\lambda}, D^{t+1}) \propto \sum_{k=1}^{K} \prod_{i=1}^{t+1} p(\bold{Y_{i}}|\boldsymbol{\lambda}, D^{i-1}, \gamma=k) p(\gamma=k),
\label{gampost}
\end{equation}
where $p(\gamma=k)$ is a discrete uniform prior defined over $(0.001,0.999)$ with $K$ categories (we comment on determining the dimension of $K$ in our simulation studies). To incorporate the learning of (\ref{gampost}) at the end of step 4 of our PL algorithm above, we first estimate the discrete posterior distribution of $\gamma$ using the Monte Carlo average of the updated samples of $\lambda_{1}, \ldots, \lambda_{J}$ at time $t+1$. Then, we resample particles from this distribution to update $f(.)$ in step 3 at time $t+2$.   
 
\section{Numerical Examples}
To illustrate our MPSB model and the associated estimation algorithms, we consider several simulation studies and an actual data on consumer demand for two households. The consumer demand data we were given access to is a subset of a large set used in \cite{B13}. The data as well as the R code are available upon request via email from the authors. 
\subsection{Example: Calibration study}
First, we present the results of several simulated studies. We constructed 10 simulated sets from the data generating process of the MPSB given by (\ref{multimod}) and (\ref{multimod2}). Each sequence of counts sampled from the model are realizations from the underlying time series model  with varying pairwise sample correlations among individual series. The parameter values are unchanged but each simulated set behaves differently as the random common environment differs drastically across simulations even for the same values of the static parameters. 

To initialize the simulations, we set $\theta_{0} \sim G(\alpha_{0}=10,\beta_{0}=10)$ representing the initial status of the random common environment. We explicitly assume that the random common environment is initialized around the unit scale (with mean $\alpha_{0}/\beta_{0}=1$). In doing so, one obtains a better understanding of the scale of the static parameters, $\lambda_{j}$'s, as a function of actual count data. This is especially important when dealing with real count data when specifying the hyper-parameters of priors for $\theta_{0}$ and the $\lambda_{j}$'s which we discuss in the sequel. We assumed that $J=5$, and the static parameters, $\lambda_{j}$'s, were 2, 2.5, 3, 3.5, and 4, respectively. The values are close to each other to investigate if the model can distinguish these static parameters. Finally, the common discount parameter, $\gamma$ was set at 0.30.   

\begin{table}
\begin{center}
\begin{small}
\begin{tabular}{rrrrrr}
 & Sim \#1 & Sim \#2& Sim \#3& Sim \#4& Sim \#5\\
\hline
\hline 
$\lambda_{1}$ & 2.06 (1.83;2.29) & 2.05 (1.83;2.25)& 1.70 (1.40;2.05)& 1.66 (1.42;1.93)& 1.91 (1.73;2.10)\\
$\lambda_{2}$ & 2.32 (2.09;2.57) & 2.64 (2.39;2.87)& 2.22 (1.85;2.61)& 2.20 (1.92;2.51)& 2.52 (2.31;2.75) \\
$\lambda_{3}$ & 2.69 (2.43;2.97) & 2.79 (2.54;3.05)& 2.75 (2.37;3.19)& 2.38 (2.11;2.70)& 2.99 (2.76;3.24) \\
$\lambda_{4}$ & 2.97 (2.70;3.24) & 3.54 (3.27;3.86)& 2.95 (2.53;3.39)& 2.58 (2.27;2.93)& 3.42 (3.17;3.67) \\
$\lambda_{5}$ & 3.54 (3.24;3.85) & 3.80 (3.52;4.08)& 3.57 (3.08;4.02)& 3.19 (2.82;3.55)& 3.72 (3.47;3.99) \\
$\gamma$ & 0.19 (0.14:0.32) & 0.20 (0.12;0.32)& 0.37 (0.22;0.50)& 0.25 (0.15;0.43)& 0.26 (0.15;0.43) \\
\hline 
 & Sim \#6 & Sim \#7& Sim \#8& Sim \#9& Sim \#10\\
 \hline
\hline 
$\lambda_{1}$ & 2.13 (1.83;2.46) & 1.96 (1.67;2.26)& 2.25 (1.91;2.59)& 2.17 (1.94;2.39)& 2.01 (1.81;2.21)\\
$\lambda_{2}$ & 2.81 (2.47;3.19) & 2.40 (2.10;2.71)& 2.50 (2.17;2.87)& 2.48 (2.25;2.77)& 2.67 (2.43;2.92)\\
$\lambda_{3}$ & 3.01 (2.64;3.37) & 2.81 (2.47;3.17)& 2.97 (2.59;3.35)& 2.72 (2.47;2.98)& 3.05 (2.76;3.30)\\
$\lambda_{4}$ & 3.57 (3.18;3.98) & 3.26 (2.88;3.65)& 3.38 (3.00;3.83)& 3.36 (3.08;3.66)& 3.25 (2.99;3.53)\\
$\lambda_{5}$ & 4.29 (3.87;4.75) & 3.75 (3.35;4.19)& 3.96 (3.53;4.44)& 3.52 (3.23;3.84)& 3.77 (2.50;4.05)\\
$\gamma$ & 0.28 (0.18;0.45) & 0.28 (0.17;0.45)& 0.26 (0.16;0.41)& 0.22 (0.14;0.33)& 0.22 (0.14;0.33)\\
\hline \hline
\end{tabular}
\end{small}
\end{center}
\caption{\textit{Posterior means and 95\% credibility intervals (in parenthesis) for static parameters across 10 simulated examples.}} \label{summary0}
\end{table}

Our PL algorithm uses N=1,000 particles. Since all simulated counts are roughly between 0 and 40 with initial values up to 5-6, we set $\theta_{0} \sim G(10,10)$ and $\lambda_{j} \sim G(2,1)$ for all $j$ (reflecting the fact that very high values of the parameter space does not make practical sense). Our numerical experiments revealed that having tighter priors especially on $\lambda_{j}$'s help identifying the true value of the parameters. Varying the hyper-parameters of the priors (within reasonable bounds with respect to the scale of the counts) does not have a significant effect on the overall fit of the models. When the priors are vague and uninformative (e.g. $G(0.001,0.001)$), our algorithm has difficulty identifying regions close to the real values of the parameters at the outset. However, in such cases the mean filtered estimates, $E(\theta_{t}\lambda_{j}|D^{t})$'s, are found to be in the near proximity of the real counts. When dealing with real data, this is not a major drawback as long as the model is able to provide reasonable filtering estimates since the true value of the static parameters will always be unknown. For practical reasons, we suggest that the initial state prior be set around the unit scale as in $\theta_{0} \sim G(10,10)$. We note here that the results were not sensitive to changes in the hyper-parameters of $\theta_{0}$ as long as its mean stayed around the region of unit scale such as those in $G(1,1), G(10,10)$ or $G(100,100)$.  

Table \ref{summary0} shows the means and 95\% credibility intervals (in parenthesis) for the estimated static parameters for 10 different simulations. For each case, the PL algorithm is able identify posterior distributions that are close to the true values of the parameters ($\lambda_{1}=2, \lambda_{2}=2.5, \lambda_{3}=3, \lambda_{4}=3.5, \lambda_{5}=4$ and $\gamma=0.3$). In addition, we also computed posterior coverage probabilities across 10 simulations by investigating if the true value of the  parameter was within the 95\% credibility bounds. (i.e. the number of times the true values of the parameter was within a given  credibility interval across 10 simulations). These coverage probabilities were estimated to be 0.9, 1.0, 0.7, 0.7 and 0.7 for the $\lambda_{j}$'s and 1.00 for $\gamma$, showing support in favor of the algorithm being able to provide coverage of the true values most of the time. 

\begin{figure}
\begin{centering}
\includegraphics[width=16cm]{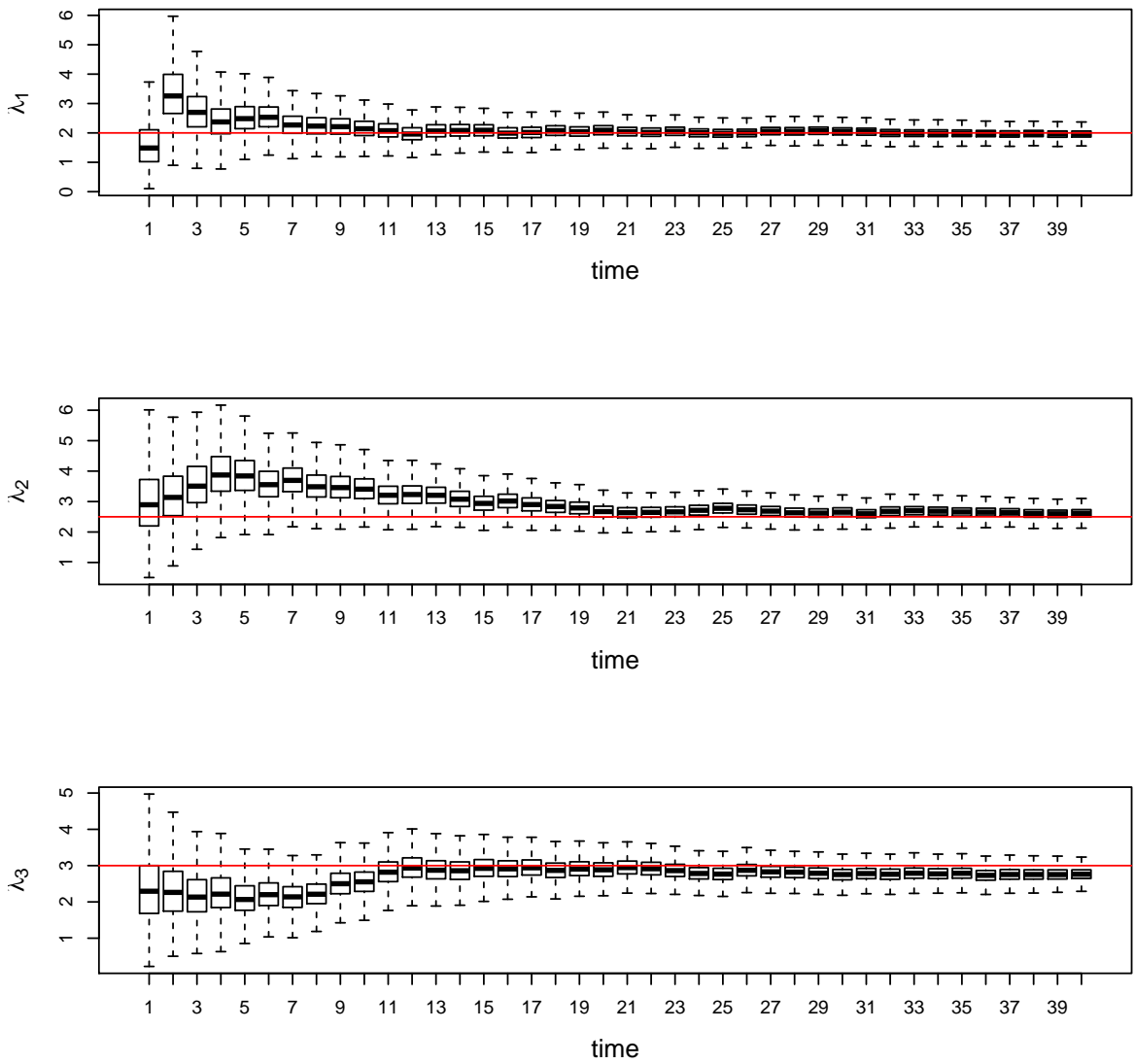}
\caption{\textit{Estimation paths of the static parameters, $\lambda_{1}, \lambda_{2}, \lambda_{3}$ over time for a given simulation (red straight line represents the real value of the parameter).}} \label{lamsimpath}
    \end{centering}
    \end{figure}

\begin{figure}
\begin{centering}
\includegraphics[width=16cm]{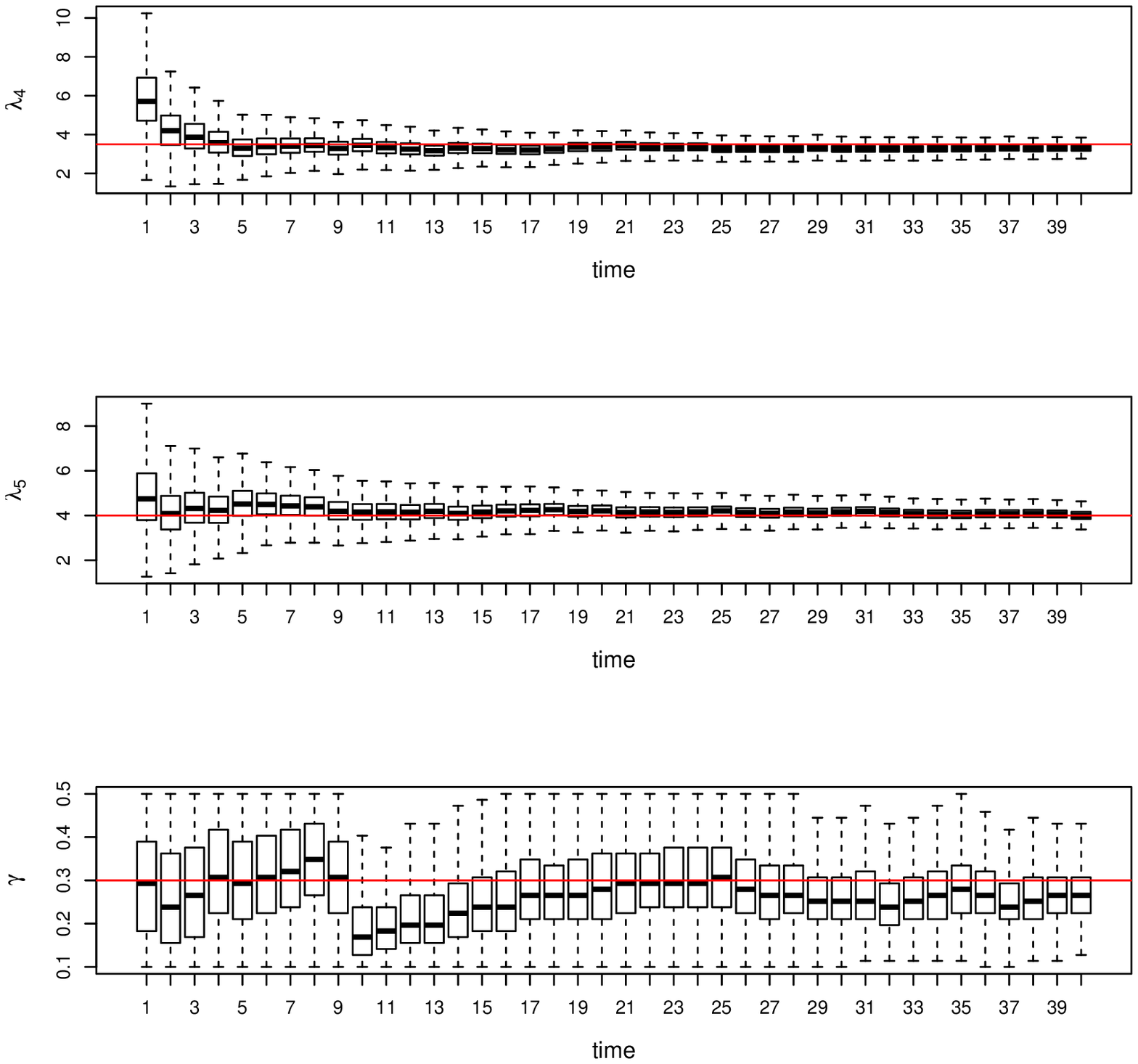}
\caption{\textit{Estimation paths of the static parameters, $\lambda_{4}, \lambda_{5}, \gamma$ over time for a given simulation (red straight line represents the real value of the parameter).}} \label{lamsimpath2}
    \end{centering}
    \end{figure}

\begin{figure}
\begin{centering}
\includegraphics[scale=.9]{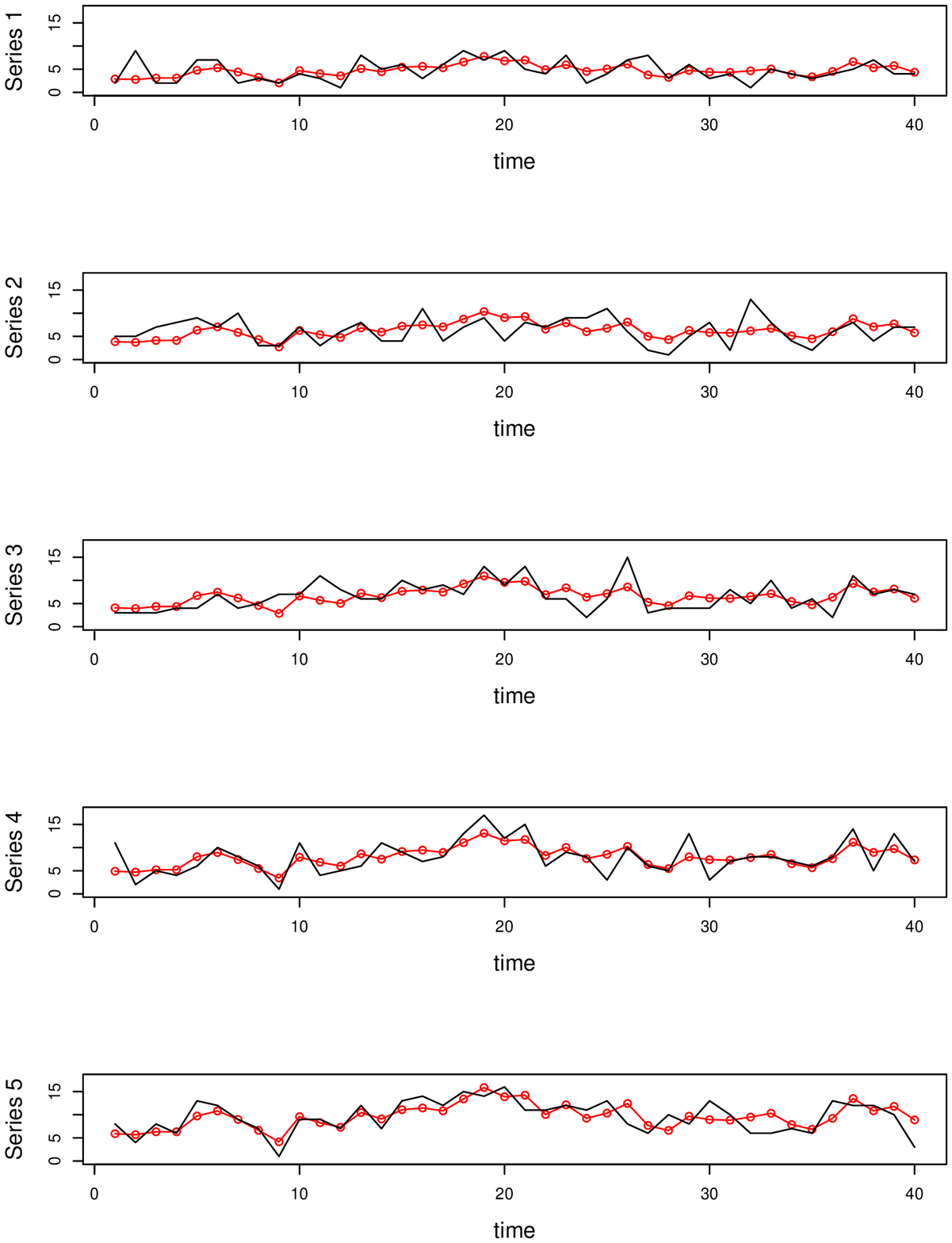}
\caption{\textit{Mean filtered states, $E(\theta_{t}\lambda_{j}|D^{t})$'s, versus the actual counts for one of the simulated examples.}} \label{fit31}
    \end{centering}
    \end{figure}

Figures \ref{lamsimpath} and \ref{lamsimpath2} show the boxplots of the estimation paths of the static parameters for one of the simulations where the straight line represents the true value of the parameter. As can be observed from the size of the boxplots, for the first few observations the posterior distributions exhibit more uncertainty. As we observe more data, the uncertainty tapers off and the posterior distributions converge to regions close to the true value of the parameters (similar plots were obtained for all 10 simulations). After observing up to 9-10 points in time, our algorithm is able to learn about the $\lambda_{j}$'s very easily, however learning of the $\gamma$ takes a few more observations. The dip in the value of $\gamma$ around time period 10 may be attributed to the jump we observe in the simulated counts in 4 our of 5 series that can be observed in Figure \ref{fit31} (from time period 9 to 10) since a lower value of $\gamma$ implies a higher correlation in our model. After a few more observations, the posterior $\gamma$ goes back to exploring regions around its true value.   

\begin{figure}
\begin{centering}
\includegraphics[totalheight=9cm]{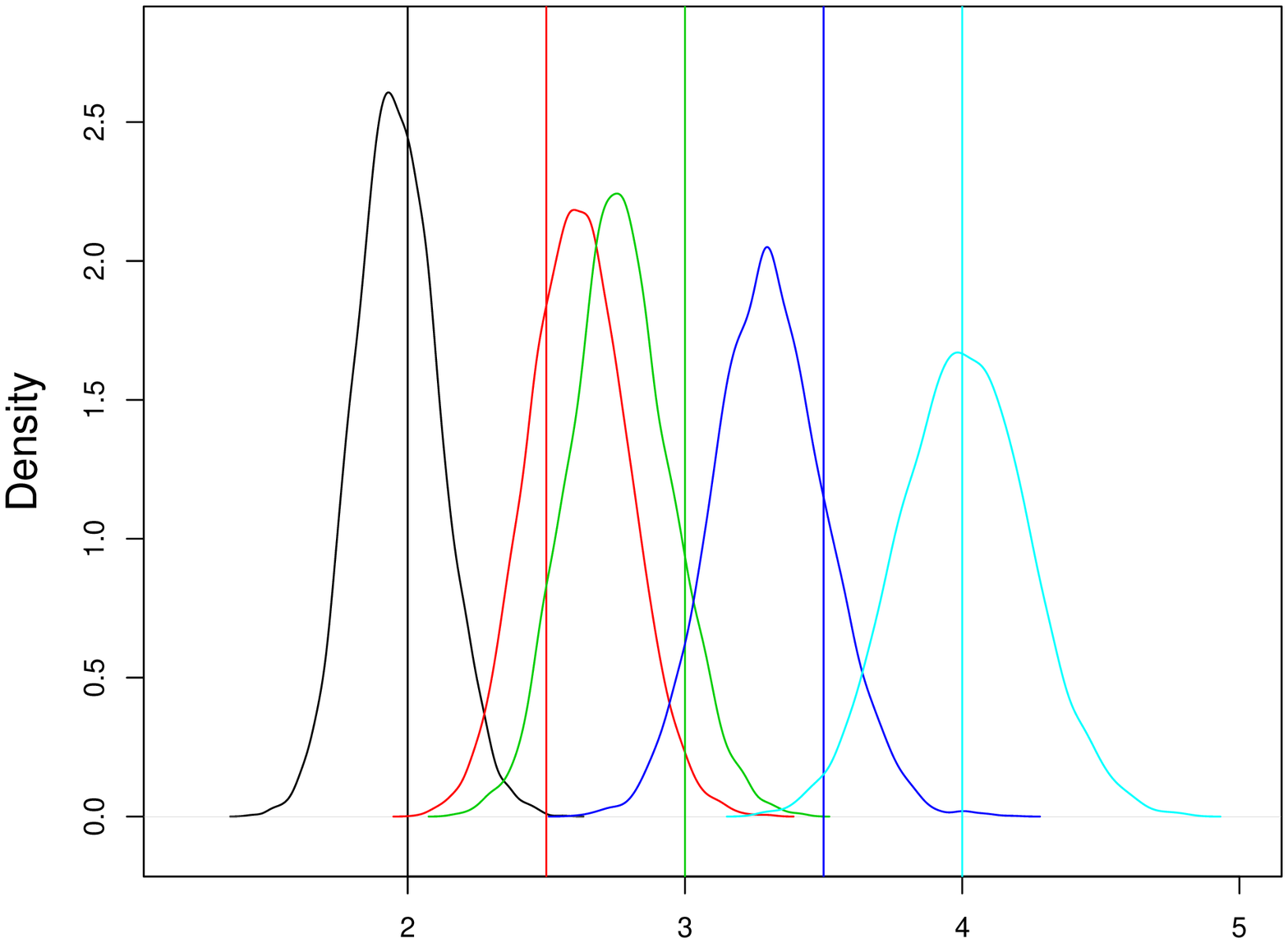}
\includegraphics[totalheight=9cm]{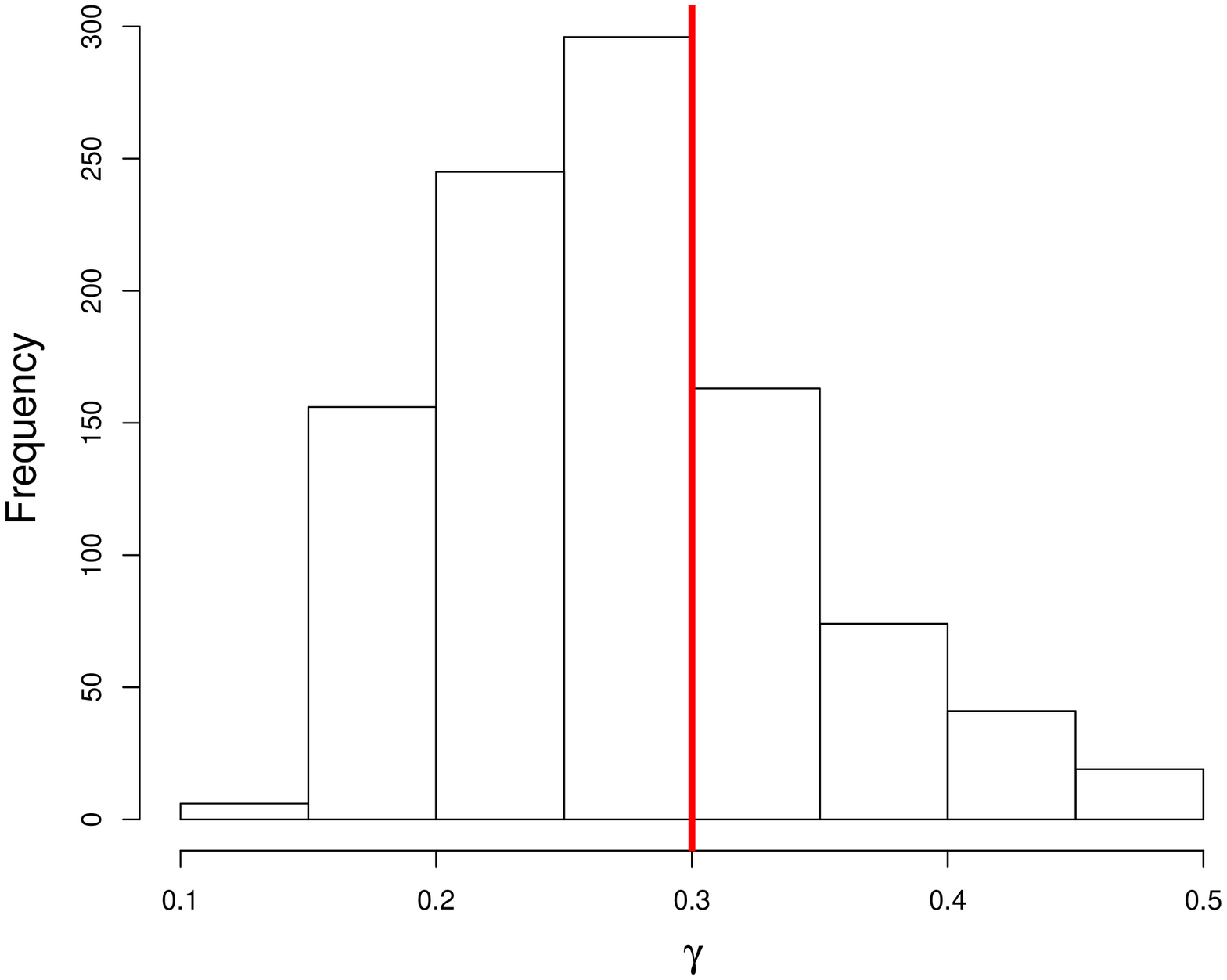}
\caption{\textit{Posterior distributions of the static parameters, $\lambda_{1}, \ldots, \lambda_{5}$ (top) and $\gamma$ (bottom), for a given simulation. Vertical lines represent the real value of the parameter.}} \label{lamsims}
    \end{centering}
    \end{figure}

The final posterior density plots of $\lambda_{1}, \ldots, \lambda_{5}$ after observing all the data are shown in the top panel of Figure \ref{lamsims} for one of the simulations. All of the density plots cover the true value of the parameter as indicated by the vertical straight lines. The posterior distribution of $\gamma$ from Figure \ref{lamsims} also shows that most of its support is close to the region of 0.30 which is the actual value of $\gamma$. The posterior mode was between 0.25 and 0.30 and the mean was estimated to be 0.27 (as there is more support on the left side of the true value in the posterior distribution). In our proposed algorithm, the estimation of $\gamma$ discussed in (\ref{gampost}) requires that we put a reasonably large value for $K$ which is the number of discrete categories for $\gamma$. For a discrete uniform prior defined over the region (0.001; 0.999), we experimented with different values for $K$ and explored cases when $K=5, 10, 30, 50, 100$ and $500$. For all 10 simulations, the posterior distributions were almost identical when $k$ was 30 or larger. For relatively smaller values of $K$ as in 5 and 10, the posterior distribution did not mix well and did not explore regions wide enough for converging to the right distribution. In cases when fast estimation is of interest, we suggest that $K$ is kept in the region of 30-40 since increasing its dimension leads to losses in estimation speed due to the fact that the negative binomial likelihood needs to be evaluated for each point in time equal to ``$K \times$ number of particles''.

Another noteworthy investigation is how good our estimated filters are with respect to actual data across simulations. To assess the model fit, we first computed the absolute percentage error (APE) for each simulation (a total of 200 observations for each simulation) and computed the median of these APEs. The results are shown in Table \ref{mape} where the estimates range between 14\% and 25\%. The reason we report the median instead of the mean APEs is the presence of some outliers which skew the results immensely. Typically the APE estimates range between 0 and 0.30 and some outliers are in the range of 3-4, which when we take the average of, show very misleading results. When we plotted the histograms of APEs for each simulation, we were able to observe that the median and the mode of the distributions were very close to each other with the means located away from these two measures due to 1-2 very high values in the right tail of the distributions. We did not report the mean squared errors (MSE) as they would not be comparable across simulations since the scale of the counts vary from one simulation to another even for the same values of the static parameters.  

\begin{table}
\begin{center}
\begin{small}
\begin{tabular}{rrrrrr}
 & Sim \#1 & Sim \#2& Sim \#3& Sim \#4& Sim \#5\\
\hline
\hline
MAPE & 0.19 & 0.14 & 0.31 & 0.22 & 0.18\\
 & Sim \#6 & Sim \#7& Sim \#8& Sim \#9& Sim \#10\\
\hline
\hline
MAPE &0.21 & 0.25 &  0.25 & 0.20 & 0.23\\
\hline
\end{tabular}
\end{small}
\end{center}
\caption{\textit{Summary of MAPEs for all simulations.}} \label{mape}
\end{table}

Figure \ref{fit31} shows the posterior means of the filtered rates, $E(\theta_{t}\lambda_{j})$'s, at each point in time versus the actual counts for a given simulated example. In this example, the series were moderately correlated with sample pairwise correlations ranging between 0.59 and 0.69. The model is able to capture most swings except for rare cases when all five series do not exhibit similar (upward/downward) patterns at a given point in time. For instance, around roughly time period 9, the counts for series 1,2,4 and 5 exhibit a drop whereas series 3 shows an increase. As the dependency across series is based on the random common environment idea, the filtered states around time period 9 exhibit a decay for all 5 series (not only for series 1,2,4 and 5). Such disagreements lead to extremely large APE estimates as discussed before but are usually no more than 1-2 times in a given simulated set.    
    
\begin{figure}
\begin{centering}
\includegraphics[width=14cm,height=7cm]{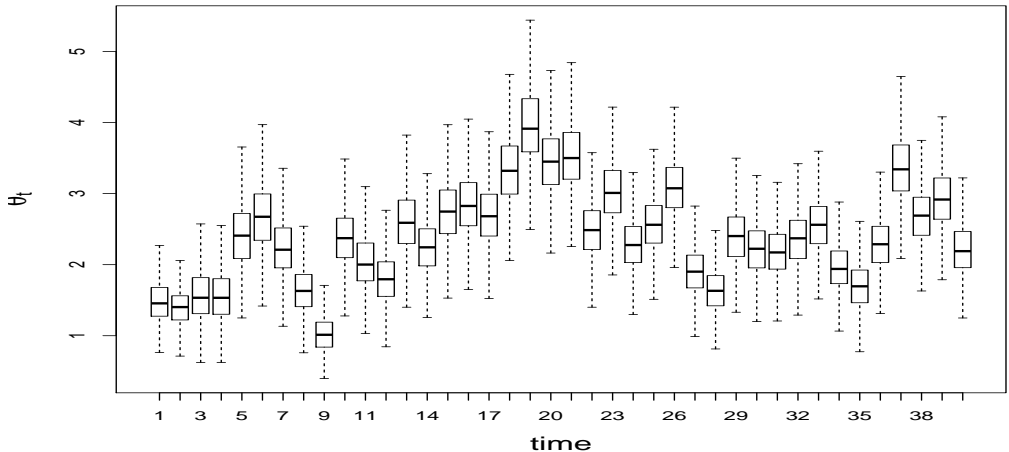}
\caption{\textit{Filtered stochastic evolution for the state of the environment, $(\theta_{t}|D^{t})$'s, over time for one of the simulated examples.}} \label{statefit1}
    \end{centering}
    \end{figure}

Figure \ref{statefit1} shows the stochastic evolution of the state of the random common environment over time that all five series have been exposed to (i.e. $p(\theta_{t}|D^{t})$ which is free of the static parameters) for a given simulation study. For instance, such a common environment could represent the economic environment financial and economic series are exposed to with swings representing local sudden changes in the market place. In our model, $\theta_{t}$s dictate the autocorrelation structure of the underlying state evolution and they induce correlations among the 5 series. The sample partial autocorrelation estimate at lag 1 for the mean of these posterior state parameters was between 0.80 and 0.90 indicating a strong first order Markovian behavior in the random common environment.   

As a final exercise, we also used the FFBS algorithm introduced in Section 2.3 to generate the full posterior joint distribution of the model parameters for each time period $t$ as in, $p(\theta_{1}, \ldots, \theta_{t}, \lambda_{1}, \ldots, \lambda_{j}|D^{t})$. As pointed out by \cite{S02}, for any MCMC based sampling method dealing with sequential estimation, the chains would need to be restarted at each point in time. In addition, issues of convergence, thinning and the size of the burn-in periods would need to be investigated. Therefore, using the FFBS algorithm would not be preferred over the PL algorithm when fast sequential estimation would be of interest as in the analysis of streaming data in web applications. To show the differences in computing speed, we estimated one of the simulated examples using both algorithms. The models were estimated on a PC running Windows 7 Professional OS with an Intel Xeon @3.2GHz CPU and 6GBs of RAM. The PL algorithm takes about 17.25 (or 58.7) seconds with 1,000 (or 5,000) particles and the FFBS algorithm takes about 270.74 seconds for 5,000 collected samples (with a thinning interval of 4) where the first 1,000 are treated as the burn-in period. In both cases, we kept $\gamma$ fixed at 0.30 even though the computational burden for its estimation with the FFBS algorithm would have been higher with ''$K \times$ Number of Samples generated=5,000'' versus ''$K \times$ Number of particles=1,000''. We also note that the estimated static parameters using the FFBS model were very close to those estimated with the PL algorithm from Table \ref{summary0}. We view the FFBS algorithm as an alternative when smoothing is of interest which can be handled in a straightforward manner as discussed in Section 2.3. For sequential filtering and prediction, we would prefer the PL algorithm due to its computational efficiency. We would like to note that the results summarized above are based on the version of our algorithm which uses the sequential importance sampling step for the state propagation instead of the rejection sampling method discussed in Step 2 of our PL algorithm. Even tough the results were identical in both cases, the computational burden for the rejection sampling algorithm was very high in some cases. Our numerical experiments revealed that the acceptance rate of the sampler became extremely small for certain values of the HGB density parameters, $a,b,c$. Therefore, unless a very efficient way of generating samples from the HGB density can be developed, we suggest the use of the extra importance sampling step in implementing our PL algorithm.    

\begin{figure}[H]
\begin{centering}
\includegraphics[width=13cm]{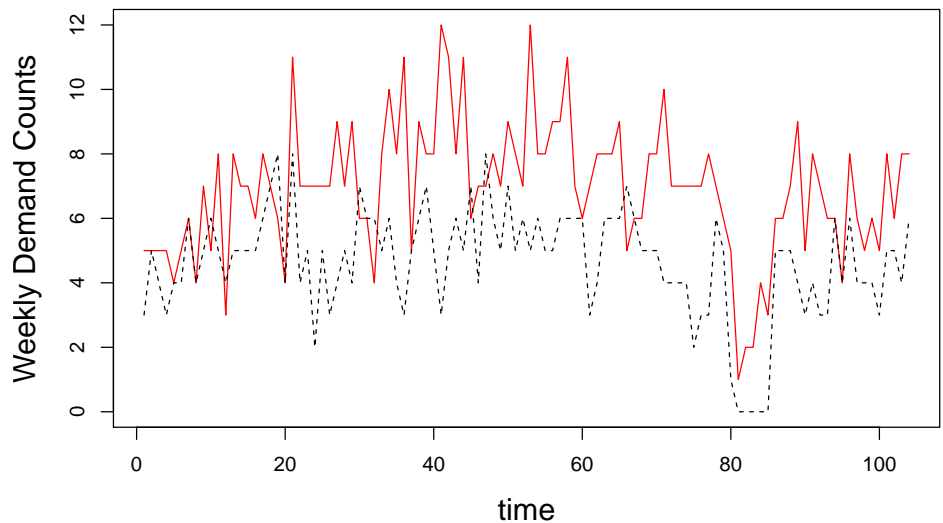}
\caption{\textit{Time series plot of weekly demand for households 1 (top straight red line) and 2 (bottom dashed black line) for 104 weeks.}} \label{fitst1}
    \end{centering}
    \end{figure}
    
\subsection{Example: Weekly Consumer Demand Data}
To show the application of our model with actual data, we used the weekly demand for consumer non-durable goods (measured by the total number of trips to the super market) of two households in the Chicago region over a period of 104 weeks (an example for a bivariate model). Therefore, in this illustration, $Y_{jt}$ for $t=1,\ldots, 104$ and $j=1,2$ are the demand of household $j$ during the time period $t$, $\theta_{t}$ represents the common economic environment that the households are exposed to at time $t$ and $\lambda_{j}$ represents the individual random effect for household $j$. The example is suitable for our proposed model since a quick empirical study of the data revealed that weekly demand of these households exhibit correlated behavior over time (temporal dependence) as well as across households (dependence from the random common environment). The sample correlation between the two series was estimated to be 0.41 which is in line with our model structure that requires positively correlated counts. In addition, the partial auto-correlation functions of both series also show significant correlations at lag 1, justifying our use of the first order Markovian evolution equation for the states. As before, we estimated the model using 1,000 particles and used similar priors. Specifically, we assumed that $\theta_{0} \sim(10,10)$ so that the initial state distribution is around the unit scale and assumed that $\lambda_{j} \sim G(2,1)$. Figure \ref{fitst1} shows the time series plot of these two series (straight red line represents household 1 and the dashed black line represents household 2) for 104 consecutive weeks.  

\begin{figure}
\begin{centering}
\includegraphics[width=15cm]{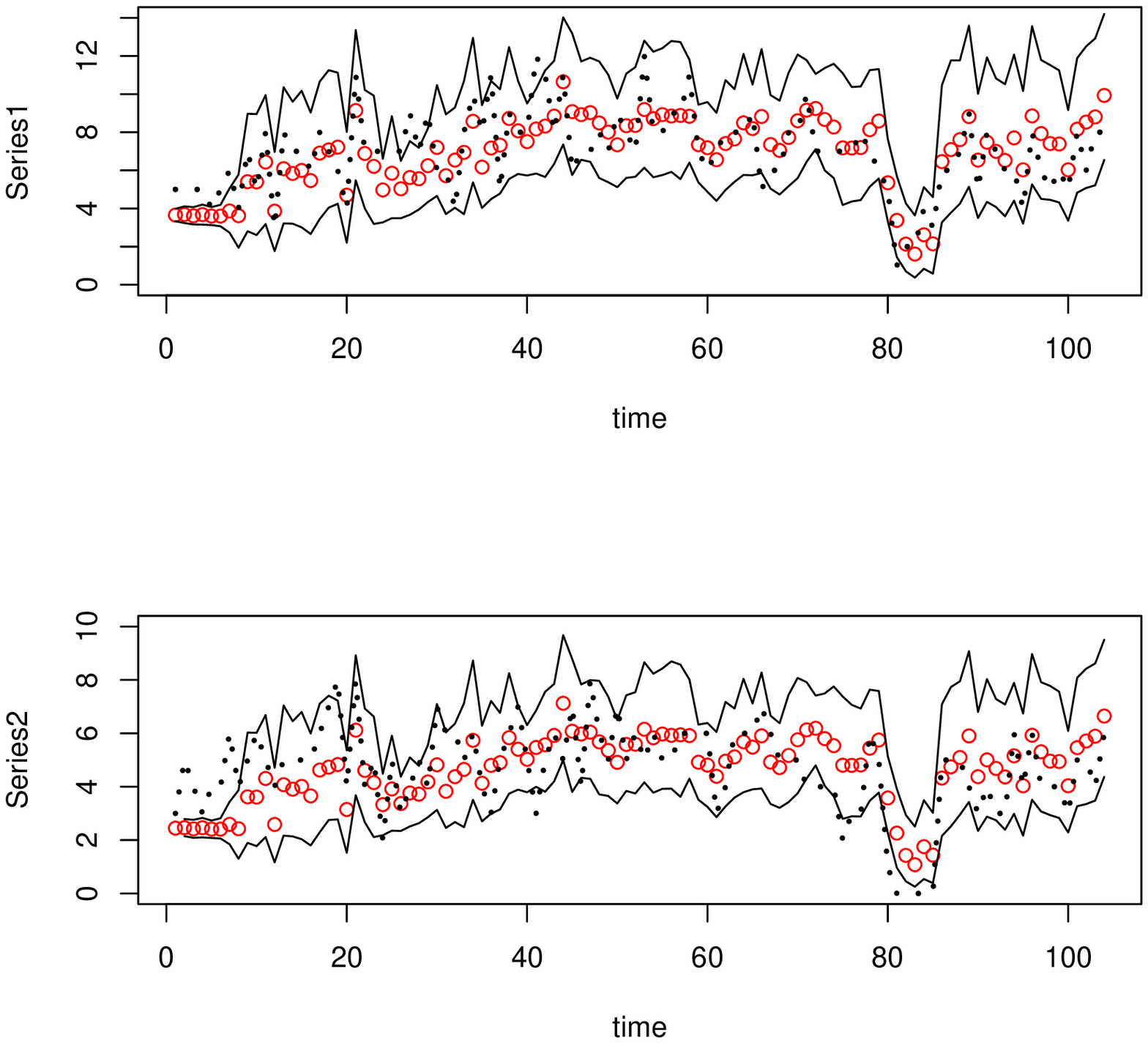}
\caption{\textit{Mean posterior estimates (red circles) and the 95\% credibility intervals (straight lines) versus the actual data (black dots) for the consumer demand data.}} \label{fitst2}
    \end{centering}
    \end{figure}
    
Figure \ref{fitst2} shows the mean posterior (filtered) estimates (red circles) and the 95\% credibility intervals (straight lines) versus the actual data (black dots). We can observe that in most cases the counts are within the credibility intervals except for the beginning first roughly ten time periods. This may be attributed to the fact that the counts for these two households were relatively lower and closer to each other initially, resulting with less global uncertainty in the counts and tighter intervals. However, visually the plots suggest that the model is able to account for sudden changes in the environment (for instance there is a sudden drop around weeks 80-85) while providing an overall reasonable fit for the counts of both households. Since the sample correlation between the two series was 0.41, suggesting a relatively low correlation, there were certain time periods when the intervals do not cover the actual data. For instance, the first 10 observations especially for series 2, look problematic and the model is slow to adapt to the sudden drop between weeks 80-85. However, approximately more than 90\% of the real counts are within the credibility interval bounds of the filtered states. Even tough we do not know the data generating process unlike the simulated examples, MAPE obtained for this example was 0.18 which is reasonably low. 
    
\begin{figure}
\begin{centering}
\includegraphics[scale=.4]{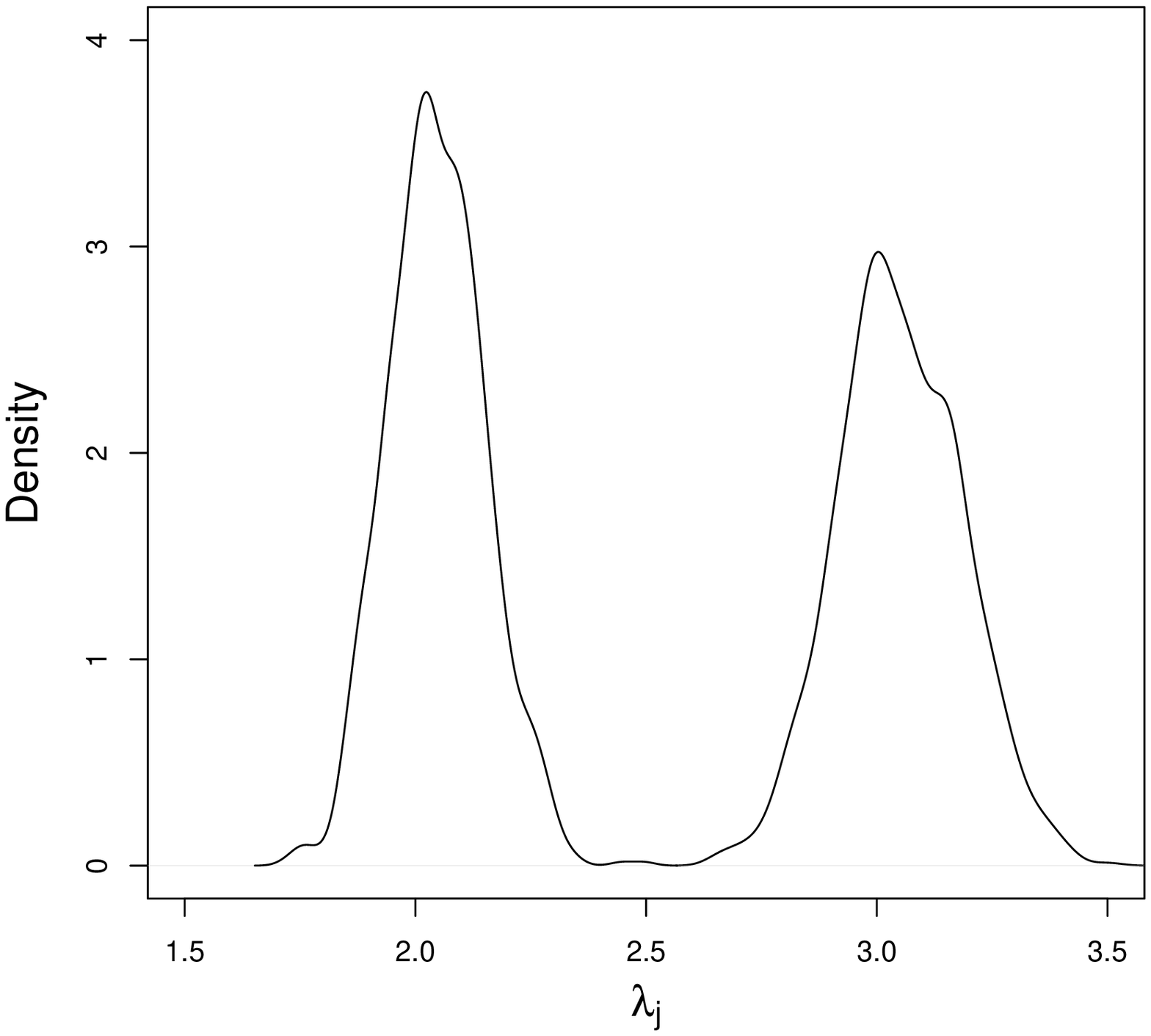} \includegraphics[scale=.4]{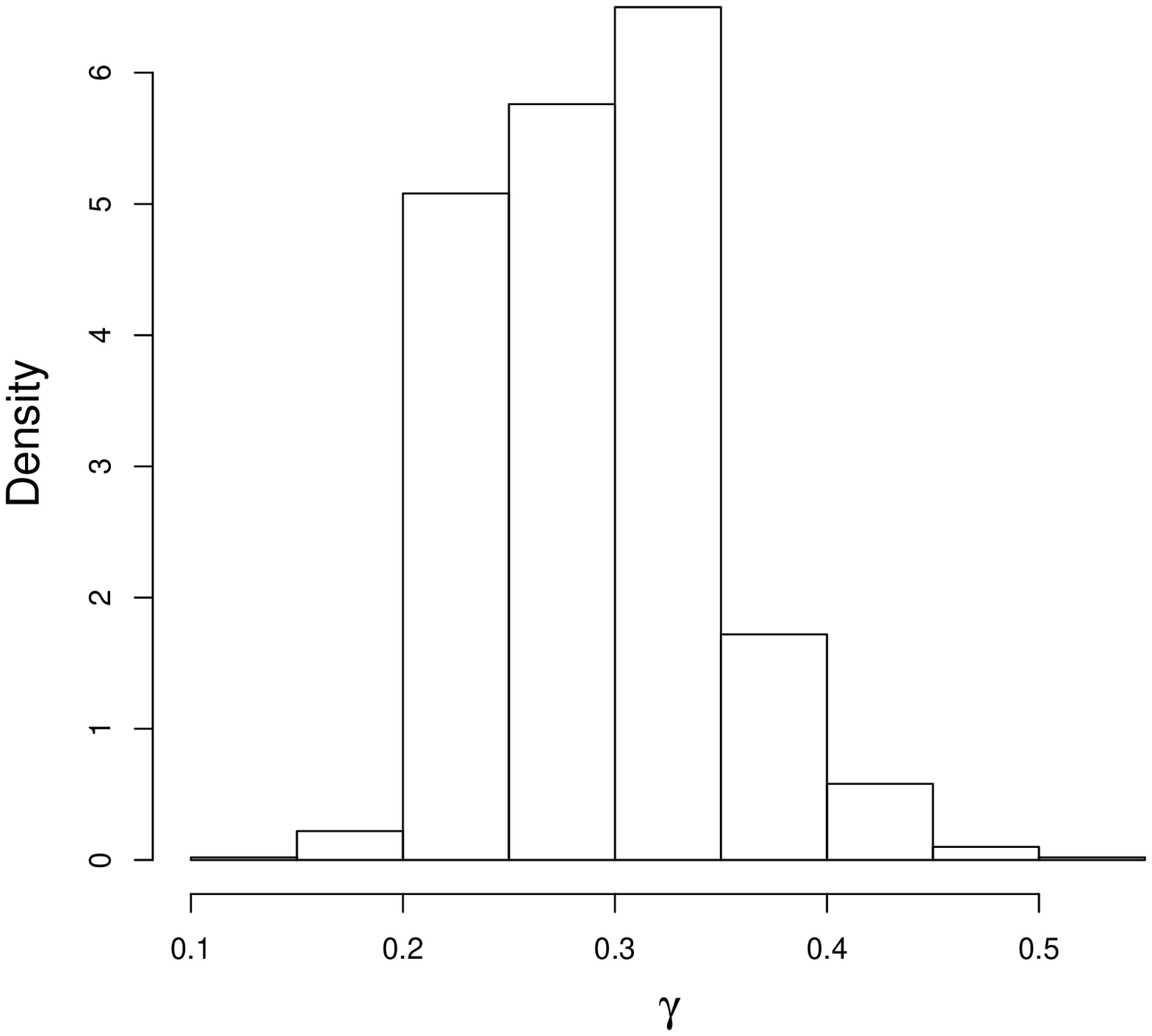}
\caption{\textit{Posterior distributions of the static parameters, $\lambda_{1}, \lambda_{2}$ (left) and $\gamma$ for the customer demand data. $\lambda_{1}$ and $\lambda_{2}$ posterior estimates indicate distinct customer demand behavior for each household.}} \label{fitst4}
    \end{centering}
    \end{figure}

The posterior distributions of $\gamma$ as well as those of $\lambda_{1}$ and $\lambda_{2}$ are given in Figure \ref{fitst4}. A higher value of $\lambda$ indicates a higher order of spending habit for household 1 as opposed to household 2 given that both are exposed to the same economic environment. The mean estimates were 3.05 and 2.04, respectively for the two static parameters. We also note that the posterior correlation between $\lambda_{1}$ and $\lambda_{2}$ was estimated to be 0.21, as expected a positive correlation a posteriori. Furthermore, the posterior mean of $\gamma$ was around 0.29. In our experience with both simulated and demand data, we observed that the posterior distribution of the static parameter $\gamma$ did not vary significantly as we observe more data points (say beyond 20-30 observations as argued previously based on Figure \ref{lamsimpath2}). Therefore, a practical approach for cases where on-line learning and forecasting is of highest importance, would be to treat $\gamma$ as fixed (either at the posterior mean or the mode) which can significantly reduce the computational burden by making filtering very fast.  

\begin{figure}
\begin{centering}
\includegraphics[height=10cm,width=13cm]{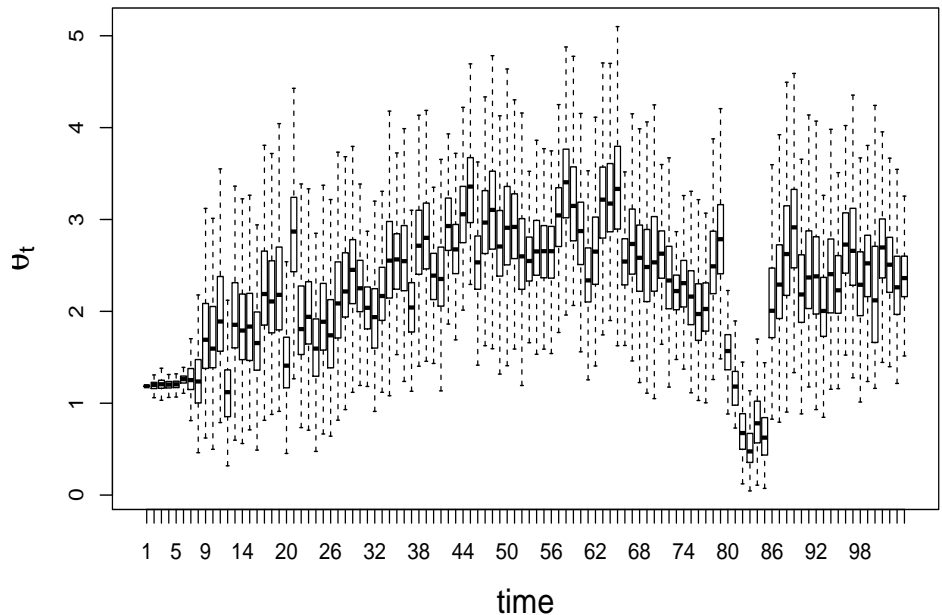}
\caption{\textit{Boxplot of the dynamic state parameters, $\theta_{t}$'s for the customer demand example, representing the random common economic environment that the two households are exposed to.}} \label{fitst3}
\end{centering}
\end{figure}

Figure \ref{fitst3} shows the boxplot of the posterior state parameters, in other words how the common environment that both households are exposed to changes over time. We can observe that the uncertainty about the environment is relatively lower at the beginning (in the first 1-5 time periods) with respect to the following time periods. This is the same observation we had drawn from the credibility intervals and could be due to the small difference between the counts. Also, the environment is said to be less favorable during roughly weeks 80-85 as there is a steep drop in the state estimates. We believe that being able to model and predict household demand would be of interest to operations managers for long term as well as short term staffing purposes. For instance, related work in queuing systems require the modeling of the time varying arrival rates that are used as inputs of a stochastic optimization formulation to determine the optimal staffing levels (see \cite{WBS07} and \cite{AS12} and the references therein for recent work using Bayesian methods for modeling Poisson arrivals in queuing models). In addition, the marketers may use these models for optimally timing the placements of advertisements and promotions. For instance, a steep drop in the state parameters (as in the weeks of 80-85 in our illustration) might lead to reductions in staffing for cutting operational costs (employees may be diverted to other tasks) or the company may decide to launch a more aggressive advertisement/promotion campaign to cope with undesirable market conditions.       
    
\section{Conclusion}
In summary, we introduced a new class of dynamic multivariate Poisson models (which we call the MPSB model) that are assumed to be exposed to the same random common environment. We considered their Bayesian sequential inference using particle learning methods for fast online updating. One of the attractive features of the PL approach as opposed to MCMC counterparts, is how fast it generates particles sequentially in the face of new data, a feature not shared with MCMC methods where the whole chain needs to be restarted when new data is observed. The model allowed us to obtain analytic forms of both the propagation density and predictive likelihood that are essential for the application of PL methods which is a property that not many state space models possess in the literature outside of Gaussian models. In addition, our model allowed us to obtain sequential updating of sufficient statistics in learning our static parameters that is another crucial and desirable feature of the PL method. Further, we showed how the proposed model leads to a new class of predictive likelihoods (marginals) for dynamic multivariate Poisson time series, which we refer to as the (dynamic) multivariate confluent hyper-geometric negative binomial distribution (MCHG-NB) and a new multivariate distribution which we call the dynamic multivariate negative binomial (DMNB) distribution. To show the implementation of our model, we considered various simulations and one actual data on weekly consumer demand for non-durable goods and discussed implications of learning both the dynamic state and static parameters. 

To conclude, we believe that it is worth noting limitations of our model. The first one is the positive correlation requirement among series as induced by (\ref{cor}). As the series are assumed to be exposed to the same random common environment, our model requires them to be positively correlated. We investigated the implications of this requirement in the estimation paths of our static  parameters in Figures \ref{lamsimpath} and \ref{lamsimpath2} and the real count data example in Figure \ref{fitst2}. Based on these plots, it is possible to infer that initially there maybe a few observations that do not follow this requirement where the static parameter estimation paths and the filtered means are not inline with their respective real values. However, if the data is overall positively  correlated, our model converges to regions around the true values of the parameters (Figures \ref{lamsimpath} and \ref{lamsimpath2}) and the mean filtered estimates are within the 95\% credibility intervals of the real counts (Figure \ref{fitst2}) after a 8-10 time periods. Another noteworthy limitation is the identifiability issue when the priors for the static parameters are uninformative. Even tough, the model keeps the product of the Poisson mean, $\theta_{t} \times \lambda_{j}$, close to the observed counts, it takes a very long time for the learning algorithm to explore regions close to the real values of the static parameters. To mitigate this issue, we suggest to use a prior centered around unity for $\theta_{0}$ and to use slightly tighter priors on $\lambda_{j}$'s as discussed in our numerical example. When dealing with real count data, we believe that this approach is reasonable as long as the posterior filtered estimates provide coverage for the true counts since we will never know the true values of the static parameters or the true data generating process.

In addition, we believe that the proposed class of models can be a fertile future area of research in developing models that can account for sparsity typically observed in multivariate count data. Our current model does not have a suitable mechanism for dealing with sparsity, however modifying the state equation to account for a transition equation that can account for sparsity maybe possible and is currently being investigated by the authors. Another possible extension would be to introduce the same approach in the general family of exponential state space models to obtain a new class of multivariate models. This is also currently being considered by the authors with encouraging results.  

\section*{Appendix A}
\subsection*{Obtaining the resampling weights of the PL algorithm in step 1}
The predictive likelihood which we denote by $p(\bold{Y_{t+1}} |z_{t})$ is required for computing the resampling weights in step 1 of our PL algorithm. Specifically, we have
\begin{align*} 
p(\bold{Y_{t+1}} |\theta_{t}, \boldsymbol{\lambda}, D^{t}) = \int p(\bold{Y_{t+1}}|\theta_{t+1},\boldsymbol{\lambda})p(\theta_{t+1}|\theta_{t},\boldsymbol{\lambda}, D^{t}) d\theta_{t+1} 
\end{align*}
where the conditional likelihood is 
\begin{equation*} 
p(\bold{Y_{t+1}}|\theta_{t+1},\boldsymbol{\lambda}) = \prod_{j} \frac{(\lambda_{j}\theta_{t+1})^{Y_{j,t+1}}}{(Y_{j,t+1})!} e^{-\lambda_{j}\theta_{t+1}}
\end{equation*}
The conditional prior (state evolution) is given by
\begin{equation*} 
p(\theta_{t+1}|\theta_{t},\boldsymbol{\lambda}, D^{t}) = \frac{\Gamma(\alpha_{t})}{\Gamma(\gamma \alpha_{t})\Gamma((1-\gamma)\alpha_{t})} \Big( \frac{\gamma}{\theta_{t}}\Big)^{\gamma\alpha_{t}} \theta_{t+1}^{\gamma \alpha_{t}-1} \Big( 1 - \frac{\gamma}{\theta_{t}}\theta_{t+1}\Big)^{(1-\gamma)\alpha_{t}-1}
\end{equation*}
Thus, rearranging the terms we can obtain $p(\bold{Y_{t+1}} |\theta_{t}, \boldsymbol{\lambda}, D^{t})$ as 
\begin{align*}
& p(\bold{Y_{t+1}} |\theta_{t}, \boldsymbol{\lambda}, D^{t}) = \prod_{j} \Bigg ( \frac{\lambda_{j}^{Y_{j,t+1}}}{Y_{j,t+1}!} \Bigg)\int_{0}^{\theta_{t} / \gamma} \frac{\Gamma(\alpha_{t})}{\Gamma(\gamma \alpha_{t})\Gamma((1-\gamma)\alpha_{t})} \Big( \frac{\gamma}{\theta_{t}}\Big)^{\gamma\alpha_{t}} \theta_{t+1}^{(\sum_{j} Y_{j,t+1})+\gamma \alpha_{t}-1} \\ & \Big( 1- \frac{\gamma}{\theta_{t}}\theta_{t+1}\Big)^{(1-\gamma)\alpha_{t}-1}  e^{-(\sum_{j}\lambda_{j})\theta_{t+1}}d\theta_{t+1}. 
\end{align*}
In the above, if use the transformation $\theta_{t+1} =\dfrac{\theta_{t}}{\gamma}u$ then we get 

\begin{align*} 
& p(\bold{Y_{t+1}} |\theta_{t}, \boldsymbol{\lambda}, D^{t}) = \prod_{j} \Bigg ( \frac{\lambda_{j}^{Y_{j,t+1}}}{Y_{j,t+1}!} \Bigg) \Bigg( \frac{\theta_{t}}{\gamma} \Bigg)^{\sum_{j} Y_{j,t+1}} \frac{\Gamma(\alpha_{t})}{\Gamma(\gamma \alpha_{t})\Gamma((1-\gamma)\alpha_{t})} \int_{0}^{1} u^{(\sum_{j} Y_{j,t+1}) +\gamma \alpha_{t}-1} \\ & (1-u)^{(1-\gamma)\alpha_{t}-1} e^{-(\sum_{j} \lambda_{j}) \frac{\theta_{t}}{\gamma}u} du, 
\end{align*}

where the term after the integral sign is similar to the hyper-geometric beta density   
\begin{equation*} 
f(x) = C x^{a-1} (1-x)^{b} e^{-cx}, 
\end{equation*}
as in \cite{G98}.
Therefore, we can write, 
\begin{equation*}
C \int u^{(\sum_{j} Y_{j,t+1}) +\gamma \alpha_{t}-1} (1-u)^{(1-\gamma)\alpha_{t}-1} e^{-(\sum_{j} \lambda_{j}) \frac{\theta_{t}}{\gamma}u} du = 1. 
\end{equation*}
Rewriting the terms we get 
\begin{equation*} 
w_{t} =\prod_{j} \Bigg ( \frac{\lambda_{j}^{Y_{j,t+1}}}{Y_{j,t+1}!} \Bigg) \Bigg( \frac{\theta_{t}}{\gamma} \Bigg)^{\sum_{j} Y_{j,t+1}} \Bigg( \frac{\Gamma(\alpha_{t})}{\Gamma(\gamma \alpha_{t})\Gamma((1-\gamma)\alpha_{t})} \Bigg)\frac{1}{C},  
\end{equation*}
where the normalization constant $C$ can be obtained as 
\begin{small}
\begin{equation*} 
\frac{1}{C} =  \Bigg( \dfrac{\Gamma(\sum_{j}Y_{j,t+1}+\gamma \alpha_{t} )\Gamma((1-\gamma)\alpha_{t})}{\Gamma(\sum_{j}Y_{j,t+1} + \alpha_{t})} \Bigg) CHF(\sum_{j}Y_{j,t+1}+\gamma \alpha_{t}; \sum_{j}Y_{j,t+1} + \alpha_{t}; -(\sum_{j} \lambda_{j}) \frac{\theta_{t}}{\gamma})
\end{equation*}
\end{small}
and CHF represents the confluent hyper-geometric function (\cite{AS68}). Therefore, the weight can be computed as  
\begin{equation*} 
w_{t} =\Bigg (\prod_{j}  \frac{\lambda_{j}^{Y_{j,t+1}}}{Y_{j,t+1}!} \Bigg) \Bigg( \frac{\theta_{t}}{\gamma} \Bigg)^{\sum_{j} Y_{j,t+1}} \Bigg( \dfrac{\Gamma(\sum_{j}Y_{j,t+1}+\gamma \alpha_{t} )\Gamma(\alpha_{t})}{\Gamma(\sum_{j}Y_{j,t+1} + \alpha_{t})\Gamma(\gamma \alpha_{t})} \Bigg)CHF(a;a+b;-c),  
\end{equation*}
where $a=\sum_{j}Y_{j,t+1}+\gamma \alpha_{t}, a+b=\sum_{j}Y_{j,t+1} + \alpha_{t}, c= (\sum_{j} \lambda_{j}) \frac{\theta_{t}}{\gamma}$. $w_{t}$ also represents the predictive likelihood (marginal) for the proposed class of dynamic multivariate Poisson models. 

\subsection*{Obtaining the propagation density of the PL algorithm in step 2}
The propagation density of the PL algorithm in step 2 can be computed as 
\begin{align*} 
p(\theta_{t+1}|\theta_{t},\boldsymbol{\lambda},\bold{Y_{t+1}},D^t) & \propto p(\bold{Y_{t+1}}|\boldsymbol{\lambda}, \theta_{t+1}) p(\theta_{t+1}| \theta_{t}, \boldsymbol{\lambda}) \\ 
& \propto \prod_{j} \Big ( (\lambda_{j}\theta_{t+1})^{Y_{j,t+1}} e^{-\lambda_{j}\theta_{t+1}} \Big ) \Big ( \theta_{t+1}^{\gamma \alpha_{t}-1} \Big( 1-  \frac{\gamma}{\theta_{t}}\theta_{t+1} \Big)^{(1-\gamma)\alpha_{t}-1}  \Big ) \\ 
& \propto \theta_{t+1}^{(\sum_{j}Y_{j,t+1}) + \gamma \alpha_{t} -1}  \Big( 1-  \frac{\gamma}{\theta_{t}}\theta_{t+1} \Big)^{(1-\gamma)\alpha_{t}-1} e^{-(\sum_{j} \lambda_{j}) \theta_{t+1}},
\end{align*}
which is proportional to a scaled hyper-geometric beta density defined over the range $(0;\frac{\theta_{t}}{\gamma})$, as $HGB(a,b,c)$, with parameters $a=(\sum_{j}Y_{j,t+1}) + \gamma \alpha_{t}, b=(1-\gamma)\alpha_{t} $ and $c=\sum_{j}  \lambda_{j}$. 
   
\section*{Appendix B}
Here, we show some of the conjugate nature of our model and show how the multivariate dynamic version was obtained starting with the univariate static case.
\subsection*{Static Univariate Case}
We start with the general rule as 
\begin{equation*}
\text{Prior} \times \text{Likelihood} = \text{Posterior} \times \text{Marginal}
\end{equation*}
Therefore, we can write  
\begin{equation*}
p(\theta) \times p(Y|\theta) = p(\theta|Y) \times p(Y),
\end{equation*}
where we assume that $\theta$ is gamma, $(Y|\theta)$ is Poisson and $Y$ is negative binomial. Thus, we can see the form as  
\begin{small}
\begin{equation*}
\Big (\frac{b^a}{\Gamma(a)} \theta^{a-1} e^{-b\theta} \Big ) \Big (\frac{\theta^{y}}{y!} e^{-\theta} \Big )= \frac{(b+1)^{(a+y)}}{\Gamma(a+y)} \theta^{a +y -1} e^{-(b+1)\theta} \binom{a+y-1}{y}\Big (\frac{b}{b+1} \Big)^{a} \Big(\frac{1}{b+1} \Big)^{y}
\end{equation*}
\end{small}
Recall, 
\begin{equation*}
\binom{a+y-1}{y} = \frac{(a+y-1)!}{y!(a-1)!}
\end{equation*}
\begin{equation*}
\Gamma(x) = (x-1)!
\end{equation*}
Also, note that the marginal model is negative binomial 
\begin{equation*}
(y|r,p) \sim NB(r,p),
\end{equation*}
where $r=a$ and $p = \frac{1}{b +1}$.
\subsection*{Dynamic Univariate Case}
This is the version considered in \cite{ASX13}, 
\begin{equation*}
p(\theta_{t+1}|D^{t}) \times p(Y_{t+1}|\theta_{t+1}) = p(\theta_{t+1}|D^{t+1}) \times p(Y_{t+1}|D^{t})
\end{equation*}
where using the same form from the above static univariate case, we can show that  
\begin{itemize}
\item The prior is $(\theta_{t+1}|D^{t}) \sim Gamma(\gamma\alpha_{t}, \gamma \beta_{t})$
\item The likelihood is $(Y_{t+1}|\theta_{t+1}) \sim Pois(\theta_{t+1})$ 
\item The posterior (filtering density) is $(\theta_{t+1}|D^{t+1}) \sim Gamma(\alpha_{t+1}, \beta_{t+1})$ with $\alpha_{t+1} = \gamma\alpha_{t}+Y_{t+1}$ and $\beta_{t+1}=\gamma \beta_{t} + 1$
\item The marginal (predictive density) is $(Y_{t+1}|D^{t}) \sim NB(r_{t+1}, p_{t+1})$ with $r_{t+1}=\gamma\alpha_{t}$ and $p_{t+1} = \dfrac{1}{\gamma \beta_{t} +1}$
\end{itemize}

\subsection*{Multivariate Dynamic Case (free of the conditioning on $\theta_{t})$}
Our multivariate model has been obtained by extending the dynamic univariate case by conditioning on series specific static parameters, $\boldsymbol{\lambda}$ and by extending the likelihood to $j$ conditionally independent Poisson densities as 
\begin{equation*}
p(\theta_{t+1}|D^{t}, \boldsymbol{\lambda}) \times p(\bold{Y_{t+1}}|\theta_{t+1},\boldsymbol{\lambda} ) = p(\theta_{t+1}|D^{t+1}, \boldsymbol{\lambda}) \times p(\bold{Y_{t+1}}|D^{t},\boldsymbol{\lambda})
\end{equation*}
where $\boldsymbol{\lambda} = \lbrace \lambda_{1},\ldots,\lambda_{J} \rbrace$ and $\bold{Y_{t+1}} = \lbrace Y_{1,t+1},\ldots, Y_{J,t+1} \rbrace$. 
Similarly, we can show that  
\begin{itemize}
\item The prior is $(\theta_{t+1}|D^{t}, \boldsymbol{\lambda}) \sim Gamma(\gamma\alpha_{t}, \gamma \beta_{t})$
\item The likelihood is $(\bold{Y_{t+1}}|\theta_{t+1},\boldsymbol{\lambda} ) = \prod_{j} Pois(\lambda_{j}\theta_{t+1})$ 
\item The posterior (filtering density) is $(\theta_{t+1}|D^{t+1}, \boldsymbol{\lambda}) \sim Gamma(\alpha_{t+1}, \beta_{t+1})$ with $\alpha_{t+1} = \gamma\alpha_{t}+ \sum_{j}Y_{j,t+1}$ and $\beta_{t+1}=\gamma \beta_{t} + \sum_{j} \lambda_{j}$
\item The marginal (predictive density) is $(\bold{Y_{t+1}}|D^{t},\boldsymbol{\lambda}) \sim DMNB(r_{t}, p_{t})$ with $r_{t}=\gamma\alpha_{t}$ and $p_{t} = \dfrac{1}{\gamma \beta_{t-1} +\sum_{j} \lambda_{j}}$, where DMNB stands for multivariate negative binomial distribution. 
\end{itemize}

\subsection*{Multivariate Case (with conditioning on $\theta_{t})$}
The form presented above would be suitable in the case where MCMC methods are used for estimation. In order to obtain the distributions required for the PL algorithm, we need to add an additional conditioning argument on $\theta_{t}$ (the state parameter from the previous period). Therefore, we extend the Bayes' rule to include $\theta_{t}$ as 
  
\begin{equation*}
p(\theta_{t+1}|\theta_{t}, D^{t}, \boldsymbol{\lambda}) \times p(\bold{Y_{t+1}}|\theta_{t+1},\boldsymbol{\lambda} ) = p(\theta_{t+1}|\theta_{t}, D^{t+1}, \boldsymbol{\lambda}) \times p(\bold{Y_{t+1}}|\theta_{t}, D^{t},\boldsymbol{\lambda}),
\end{equation*}
based on which we can show that the conditional prior is 
$$(\theta_{t+1}|\theta_{t}, D^{t}, \boldsymbol{\lambda}) \sim ScaledBeta(\gamma\alpha_{t}, (1-\gamma) \alpha_{t})\:  \text{defined over} \: \Big (0;\dfrac{\theta_{t}}{\gamma} \Big)$$
The likelihood is 
$$(\bold{Y_{t+1}}|\theta_{t+1},\boldsymbol{\lambda} ) = \prod_{j} Pois(\lambda_{j}\theta_{t+1})$$ 
The conditional posterior (propagation density) is a scaled HGB and is 
$$(\theta_{t+1}|\theta_{t}, D^{t+1}, \boldsymbol{\lambda}) \sim HGB[(\sum_{j}Y_{j,t+1}) + \gamma \alpha_{t},(1-\gamma)\alpha_{t},\sum_{j}\lambda_{j}] \:  \text{defined over} \: \Big (0;\dfrac{\theta_{t}}{\gamma} \Big),$$
where HGB stands for the hyper-geometric beta distribution. The predictive likelihood density, $(\bold{Y_{t+1}}|\theta_{t}, D^{t},\boldsymbol{\lambda})$, would be a new multivariate density as shown below. Note also the forms of the above densities as  
\begin{itemize}
\item $p(\theta_{t+1}|\theta_{t}, D^{t}, \boldsymbol{\lambda}) = \frac{\Gamma(\alpha_{t})}{\Gamma(\gamma \alpha_{t})\Gamma((1-\gamma)\alpha_{t})} \Big( \frac{\gamma}{\theta_{t}}\Big)^{\gamma\alpha_{t}} \theta_{t+1}^{\gamma \alpha_{t}-1} \Big( 1 - \frac{\gamma}{\theta_{t}}\theta_{t+1}\Big)^{(1-\gamma)\alpha_{t}-1}$
\item $p(\bold{Y_{t+1}}|\theta_{t+1},\boldsymbol{\lambda}) = \prod_{j} \frac{(\lambda_{j}\theta_{t+1})^{Y_{j,t+1}}}{(Y_{j,t+1})!} e^{-\lambda_{j}\theta_{t+1}}$
\item $(\theta_{t+1}|\theta_{t}, D^{t+1}, \boldsymbol{\lambda}) = \Big ( \frac{\gamma}{\theta_{t}}\Big)^{(\sum_{j}Y_{j,t+1}) + \gamma \alpha_{t}} \theta_{t+1}^{(\sum_{j}Y_{j,t+1}) + \gamma \alpha_{t} -1}  \Big(  1- \frac{\gamma}{\theta_{t}}\theta_{t+1} \Big)^{(1-\gamma)\alpha_{t}-1} \times$ 

$e^{-(\sum_{j} \lambda_{j}) \theta_{t+1}}  \Bigg( \frac{\Gamma(\sum_{j}Y_{j,t+1} + \alpha_{t})}{\Gamma(\sum_{j}Y_{j,t+1}+\gamma \alpha_{t} )\Gamma((1-\gamma)\alpha_{t})} \Bigg) \frac{1}{CHF(\sum_{j}Y_{j,t+1}+\gamma \alpha_{t}; \sum_{j}Y_{j,t+1} + \alpha_{t}; -(\sum_{j} \lambda_{j})\frac{\theta_{t}}{\gamma})}$, 
\end{itemize}
where CHF represents the confluent hyper-geometric function. Therefore, we can show that $(\bold{Y_{t+1}}|\theta_{t}, D^{t},\boldsymbol{\lambda})$ would have the following form 
\begin{small}
\begin{equation*}
p(\bold{Y_{t+1}}|\theta_{t}, D^{t},\boldsymbol{\lambda}) = \Bigg (\prod_{j}  \frac{\lambda_{j}^{Y_{j,t+1}}}{Y_{j,t+1}!} \Bigg) \Bigg( \frac{\theta_{t}}{\gamma} \Bigg)^{\sum_{j} Y_{j,t+1}} \Bigg( \dfrac{\Gamma(\sum_{j}Y_{j,t+1}+\gamma \alpha_{t} )\Gamma(\alpha_{t})}{\Gamma(\sum_{j}Y_{j,t+1} + \alpha_{t})\Gamma(\gamma \alpha_{t})} \Bigg)CHF(a;a+b;-c),  
\label{Multiweight}
\end{equation*}
\end{small}
where $a=\sum_{j}Y_{j,t+1}+\gamma \alpha_{t}, a+b=\sum_{j}Y_{j,t+1} + \alpha_{t}, c= (\sum_{j} \lambda_{j}) \frac{\theta_{t}}{\gamma}$. We refer to the above distribution as the multivariate confluent hyper-geometric negative binomial (MCHG-NB) distribution. The MCHG-NB density has the same form as the resampling weight obtained in (\ref{weight}) for our PL algorithm. 
\bibliography{Poissonbiblio}
\bibliographystyle{apa}
\end{document}